\documentclass[11pt,british]{article}

\usepackage[T1]{fontenc}
\usepackage[latin9]{inputenc}
\usepackage[a4paper]{geometry}
\usepackage{color}
\usepackage{babel}
\usepackage{mathrsfs}
\usepackage{amsmath}
\usepackage{amsthm}
\usepackage{amssymb}
\usepackage{graphicx}
\usepackage{setspace}
\usepackage[authoryear]{natbib}
\usepackage[unicode=true,pdfusetitle,
 bookmarks=true,bookmarksnumbered=false,bookmarksopen=false,
 breaklinks=false,pdfborder={0 0 1},backref=false,colorlinks=false]
 {hyperref}

\makeatletter
\theoremstyle{remark}
\newtheorem{rem}{\protect\remarkname}
\theoremstyle{plain}
\newtheorem{prop}{\protect\propositionname}
\theoremstyle{plain}
\newtheorem{lem}{\protect\lemmaname}
\theoremstyle{plain}
\newtheorem{thm}{\protect\theoremname}
\theoremstyle{plain}
\newtheorem{cor}{\protect\corollaryname}
\theoremstyle{definition}
 \newtheorem{example}{\protect\examplename}
\theoremstyle{plain}
\newtheorem{assumption}{\protect\assumptionname}

\usepackage{lscape}
\usepackage{changepage}
\usepackage{geometry}
\setcitestyle{open={(},close={)}}

\makeatother

\providecommand{\assumptionname}{Assumption}
\providecommand{\corollaryname}{Corollary}
\providecommand{\examplename}{Example}
\providecommand{\lemmaname}{Lemma}
\providecommand{\propositionname}{Proposition}
\providecommand{\remarkname}{Remark}
\providecommand{\theoremname}{Theorem}

\begin{document}
\title{Identifiability and Estimation of Possibly Non-Invertible SVARMA Models:
A New Parametrisation}
\author{Bernd Funovits}

\maketitle
\thispagestyle{empty}

\textbf{Proposed Running Head: }Possibly Non-Invertible SVARMA in
WHF

\subsection*{Affiliation}

\begin{singlespace}
\textbf{University of Helsinki}

Faculty of Social Sciences

Discipline of Economics

P. O. Box 17 (Arkadiankatu7)

FIN-00014 University of Helsinki
\end{singlespace}

and

\begin{singlespace}
\textbf{TU Wien}

Institute of Statistics and Mathematical Methods in Economics

Econometrics and System Theory

Wiedner Hauptstr. 8

A-1040 Vienna
\end{singlespace}

\textbf{E-mail: }bernd.funovits@helsinki.fi

\pagebreak{}

\thispagestyle{empty}

\section*{Abstract}

This article deals with parameterisation, identifiability, and maximum
likelihood (ML) estimation of possibly non-invertible structural vector
autoregressive moving average (SVARMA) models driven by independent
and non-Gaussian shocks. In contrast to previous literature, the novel
representation of the MA polynomial matrix using the Wiener-Hopf factorisation
(WHF) focuses on the multivariate nature of the model, generates insights
into its structure, and uses this structure for devising optimisation
algorithms. In particular, it allows to parameterise the location
of determinantal zeros inside and outside the unit circle, and it
allows for MA zeros at zero, which can be interpreted as informational
delays. This is highly relevant for data-driven evaluation of Dynamic
Stochastic General Equilibrium (DSGE) models. Typically imposed
identifying restrictions on the shock transmission matrix as well
as on the determinantal root location are made testable. Furthermore,
we provide low level conditions for asymptotic normality of the ML
estimator and analytic expressions for the score and the information
matrix. As application, we estimate the Blanchard and Quah model and
show that our method provides further insights regarding non-invertibility
using a standard macroeconometric model. These and further analyses
are implemented in a well documented R-package.

Keywords: Non-invertibility, Wiener-Hopf factorisation, structural
vector autoregressive moving-average models, non-Gaussianity, identifiability

JEL classification: C32, C51, E52

\pagebreak{}

\section{Introduction}

\setcounter{page}{1}

Tracing out the response of variables of interest with respect to
underlying economic shocks is part of almost every macroeconometric
analysis \citep{Sims80}. The main tool for generating this so-called
impulse response function (IRF) is the structural vector autoregressive
(SVAR) model. In this article, we suggest a superior alternative to
SVAR models which allows for modelling the fact the economic agents
have more information available than an outside observer (like an
econometrician): possibly non-invertible SVARMA models.

It is well-known and easily seen from the spectral factorisation theorem
for rational spectral densities \citep[page 47]{Rozanov67}, \citep[page 66]{Hannan70},
that neither the zero location (inside or outside the unit circle)
nor the static shock transmission matrix is identifed from second
moment information. Focusing on the static identifiability problem
(i.e. assuming implicitly that the IRF has no zeros inside the unit
circle), an immense body of literature has been dedicated to devising
(mainly story-driven) identification strategies for the static shock
transmission matrix in SVARs \citep[Chapter 4]{KilianLut17}. Recently,
\citet{LMS_svarIdent16} and \citet{GourierouxZakoianRenne17} have
shown that structural vector autoregressive (SVAR) models driven by
independent non-Gaussian components are identified up to scaling and
permutations which makes the typically imposed identifying restrictions
testable. It is thus possible to employ a data-driven approach instead
of a story-telling approach for identifying the static shock transmission
matrix. 

While the literature on SVAR models is abundant, contributions regarding
possibly non-invertible SVARMA models are scarce (and will be discussed
below), possibly due to the fact that more sophisticated mathematical
tools are required for their analysis. Be that as it may, the fact
that SVAR models exclude a priori the existence of determinantal MA
roots\footnote{Even though the natural approach to overcome this deficiency is to
allow for a non-constant MA polynomial matrix with zeros inside the
unit circle, there are approaches which try to recover a single shock
from a SVAR models and assess the ``degree of non-invertibility''
\citep{simsZha06monpol_recessioni,forniGambetSala19_noninv}.} is especially problematic in structural economic environments where
economic agents have more information available than outside observers
\citep{Alessi11noninv}, corresponding to determinantal MA roots inside
the unit circle \citep[page 86]{HansenSargent91_2difficulties}.

Even without allowing for determinantal roots of the MA polynomial
matrix inside the unit circle, it is known that complex dynamics are
better approximated and described by SVARMA models \citep{HannanDeistler12,PoskittYao17}.
In macroeconometrics especially, where data is sometimes only available
at quarterly instances, it is of paramount importance to use parsimoniously
parameterised models (like e.g. SVARMA models) for which the IRF and
variance decompositions can be obtained straight-forwardly. \citet{Poskitt16},
\citet{PoskittYao17}, \citet{RaghavanAthSilvapulle16}, \citet{AthVahid08},
and \citet{AthVahid_JTSA_08} provide ample evidence and make a strong
point for using VARMA models instead of VAR models for econometric
analysis.

We focus here on a general treatment of possibly non-invertible SVARMA
models, provide a new parametrisation for the MA polynomial, show
that this parametrisation is identifiable under different non-Gaussianity
assumptions and (just-identifying) parameter restrictions which are
necessary to make the WHF unique, and provide low-level conditions
on the true shock densities such that the ML estimator is asymptotically
normal. Moreover, we characterise the classes of observational equivalence
in terms of second moment information at different stages of the modelling
process, i.e. from rational spectral density to spectral factors (or
equivalently the IRF), from spectral factor to AR and MA polynomial
and static shock transmission matrix, and finally from the MA matrix
polynomial to the (without further assumptions in general non-unique)
WHF factorisation. 

To illustrate the importance of identifying the root location correctly,
consider the example given in \citet{GourierouxMR_svarma19} who refer
to \citet{lippi_reichlin93aer}. Notice that when the true model for
productivity is given as $y_{t}=\varepsilon_{t}+b\varepsilon_{t-1}$,
where $\left(\varepsilon_{t}\right)$ is an i.i.d. shock to productivity
with variance equal to one, and such that the largest impact of a
productivity shock is delayed, i.e. $b>1$, we cannot reconstruct
these shocks from present and past observed data (thence the term
``non-invertibility''). Moreover, it is easy to see that the process
$x_{t}=\eta_{t}+\frac{1}{b}\eta_{t-1}$ where $\left(\eta_{t}\right)$
is a white noise process with variance $b^{2}$ has the same autocovariance
function (and spectral density) as $\left(y_{t}\right)$.

Next, we discuss two related approaches based on Blaschke matrices.
The recent contribution \citet{GourierouxMR_svarma19} provides an
overview of estimation strategies (essentially in the bivariate case
with one MA lag) and applications in macroeconomics and finance. Like
previous articles in the literature on possibly non-causal or possibly
non-invertible multivariate time series models \citep{davis10,LanneSaikkonen13},
they apply results by \citet{chanho04,chanHoTong06} with regards
to unique representation of multivariate linear processes to obtain
identifiability of the possibly non-invertible SVARMA model. The approach
is, however, seriously flawed in several dimensions, see \citet{funovits2020gmr_comment}
for a detailed analysis. Firstly, they use Blaschke matrices to mirror
initial estimates based on moment estimation at the unit circle without
taking the possibly complex nature of the mirroring procedure into
account and thus leave the real-valued parameter space. They discard
imaginary parts or replace complex-valued matrices with identity matrices
without providing any justification for doing so. Moreover, their
implementation\footnote{Downloaded from the website of the Review of Economic Studies: \href{https://doi.org/10.1093/restud/rdz028}{https://doi.org/10.1093/restud/rdz028}.}
ignores the complex-valued nature in the sense that whenever there
is a complex-conjugated eigenvalue to be mirrored, the obtained ``Blaschke
matrix'' does not have a spectrum equal to the identity matrix. It
is non-trivial to prove that the Blaschke matrix mirroring a pair
of complex-conjugated roots at the unit circle has real-valued coefficients.
This problem is solved in \citet{scherrer_funovits2020allpass}. Secondly,
their article is presented in a way that suggests that it is applicable
to VARMA(p,q) models of arbitrary cross-sectional dimension $n$ and
MA order $q$. However, unless the product $n\cdot q$ of the cross-sectional
dimension $n$ and the MA order $q$ is sufficiently small, their
approach is not computationally feasible since they need to perform
up to $2^{nq}$ optimisations. Lastly, \citet{GourierouxMR_svarma19}
only state high-level conditions in their Proposition 2 on asymptotic
properties without verifying them. In particular, their statement
that the parameter space for VARMA models be compact is incorrect
thus their Proposition 2 is empty.

The working paper \citet{velasco2020identification} is the multivariate
generalisation of \citet{VelascoLobato18} which treats univariate
possibly non-causal and possibly non-invertible ARMA models using
an objective function based on higher order spectra. The array representation
of multivariate higher order spectral densities seems to be based
on \citet[Theorem 2.8.1, page 34]{Brillinger75} and \citet{Jammalamadaka_rao_terdik06}
and is the logical step for generalising the objective function for
univariate models to the one for multivariate models. While Velasco's
approach certainly has its merits, similar criticism as mentioned
above applies. The estimation requires optimisation of $2^{nq}$ basic
representations and is therefore only suitable for models of small
dimensions and/or small MA order. Moreover, evaluation of higher order
cumulants is computationally extremely costly, even after taking symmetries
into account\footnote{The evaluation of the higher order periodogram of order $4$ requires
$\mathcal{O}\left(T^{3}\right)$ evaluations, where $T$ is the number
of observations.} which is prohibitive for applications outside usual sample sizes
of macroeconomic low frequency data. As described on page 29 in \citet{velasco2020identification},
these basic representations only serve as preliminary estimates for
local GMM which again can hardly be made computationally feasible.
Another possible problem regarding the use of higher order cumulants
directly is the fact that methods based on higher (than second) order
cumulant spectra are not efficient in estimating parameters of finite
parameter schemes\footnote{See \citet{LiiRosenblatt96}, who mention \citet{kreiss87adaptive}
as a univariate causal and invertible approach for obtaining asymptotically
efficient estimates in the ARMA case with unknown input densities,
and \citet{Gassiat90,gassiat1993} for the univariate non-causal AR
case with unknown input densities for results regarding locally asymptotic
minimax estimators.} and that they are quite sensitive with respect to outliers. In \citet[page 280]{Hyvarinen01},
it is mentioned in the context of comparing different ICA methods
that ``using kurtosis is well justified only if the ICs are subgaussian
and there are no outliers''. A similar approach has been pursued
in \citet{funo_allpass} (and presented at the NBER Time Series Meeting
2019 in Hong Kong) but discarded due to these shortcomings. Furthermore,
it is unclear how it is ensured that the real parameter space is not
left when applying possibly complex-valued Blaschke matrices\footnote{While an article by Baggio and Ferrante is cited, there is no article
of these authors in the references.}. Last but not least, \citet{velasco2020identification} assumes that
the zero-lag coefficient matrix of the MA matrix polynomial is non-singular
which excludes informational delays. To summarise, while the approach
in \citet{velasco2020identification} certainly has its merits, any
approach based on Blaschke matrices is sub-optimal dinvr non-invertibility
is only an afterthought. In contrast, non-invertibility is at the
centre when parametrising the MA polynomial matrix using the WHF.
Both \citet{GourierouxMR_svarma19} and \citet{velasco2020identification}
try to optimise an objective function over a complicated parameter
space without sufficiently analysing its structure.

This article is accompanied by an R-package\footnote{It can be downloaded from \href{https://github.com/bfunovits/}{https://github.com/bfunovits/}.}
which implements the developed methods and contains various worked
examples from the economic literature in vignettes. The package  builds
on the R packages \texttt{RLDM}\footnote{The abbreviation RLDM stands for Rational Linear Dynamic Models.}
\citep{ScherrerFuno_rldm} and \texttt{rationalmatrices} \citep{ScherrerFuno_ratmat}.

The rest of the paper is structured as follows. In Section 2, the
SVARMA model and the WHF parametrisation are introduced and the latter
is shown to be unique under certain (just-identifying) parameter restrictions.
In Section 3, the identifiability problem is analysed and the classes
of observational equivalence with respect to second moment information
are characterised. Moreover, the (static and dynamic) identifiability
result for our novel parameterisation is stated and proved, and an
identification scheme for selecting a particular signed permutation
is presented. In Section 4, the maximum likelihood (ML) estimator
is derived and shown to be consistent and asymptotically normal. In
Section 5, we estimate the \citet{blanchard_quah89} model, compare
our results to previous ones in the literature, and discuss further
advantages of our approach in a standard macroeconometric model. Detailed
analysis is contained in the associated R-package. The Appendix contains
results on zeros and poles at infinity of rational matrices, details
on the (non-) uniqueness of the WHF, and derivations regarding asymptotic
normality of the ML estimator.

We use $z$ as a complex variable as well as the backward shift operator
on a stochastic process, i.e. $z\left(y_{t}\right)_{t\in\mathbb{Z}}=\left(y_{t-1}\right)_{t\in\mathbb{Z}}$
and define $i=\sqrt{-1}$. The transpose of an $\left(m\times n\right)$-dimensional
matrix $A$ is represented by $A'$. For the sub-matrix of $A$ consisting
of rows $m_{1}$ to $m_{2}$, $0\leq m_{1}\leq m_{2}\leq m$, we write
$A_{\left[m_{1}:m_{2},\bullet\right]}$ and analogously $A_{\left[\bullet,n_{1}:n_{2}\right]}$
for the sub-matrix of $A$ consisting of columns $n_{1}$ to $n_{2}$,
$0\leq n_{1}\leq n_{2}\leq n$. The column-wise vectorisation of $A\in\mathbb{R}^{m\times n}$
is denoted by $vec\left(A\right)\in\mathbb{R}^{mn\times1}$ and for
a square matrix $B\in\mathbb{R}^{n\times n}$ we denote with $vecd{^\circ}\left(B\right)\in\mathbb{R}^{n(n-1)}$
the vectorisation where the diagonal elements of $B$ are left out.
The $n$-dimensional identity matrix is denoted by $I_{n}$, an $n$-dimensional
diagonal matrix with diagonal elements $\left(a_{1},\ldots,a_{n}\right)$
is denoted by $\text{diag}\left(a_{1},\ldots,a_{n}\right)$, and the
inequality $">0"$ means positive definiteness in the context of matrices.
The column vector $\iota_{i}$ has a one at positions $i$ and zeros
everywhere else. The expectation of a random variable with respect
to a given probability space is denoted by $\mathbb{E}\left(\cdot\right)$.
Convergence in probability and in distribution are denoted by $\xrightarrow{p}$
and $\xrightarrow{d}$, respectively. Partial derivatives $\left.\frac{\partial f(x)}{\partial x}\right|_{x=x_{0}}$
of a real-valued function $f(x)$ evaluated at a point $x_{0}\in\mathbb{R}^{k}$
are denoted by $f_{x}\left(x_{0}\right)$ and considered columns.
The normal distribution with mean $\mu\in\mathbb{R}^{n}$ and covariance
matrix $\Sigma\in\mathbb{R}^{n\times n}$ is denoted by $\mathcal{N}\left(\mu,\Sigma\right)$.

\section{\label{sec:Model}Model}

We start from an $n$-dimensional VARMA system 
\begin{equation}
\underbrace{\left(I_{n}-a_{1}z-\cdots a_{p}z^{p}\right)}_{=a(z)}y_{t}=\underbrace{\left(b_{0}+b_{1}z+\cdots+b_{q}z^{q}\right)}_{=b(z)}B\varepsilon_{t},\quad a_{i},b_{i}\in\mathbb{R}^{n\times n}.\label{eq:system}
\end{equation}
The economic shocks $\left(\varepsilon_{t}\right)_{t\in\mathbb{Z}}$
driving the system are identically and independently distributed (i.i.d.)
across time, have zero mean, and diagonal covariance matrix $\Sigma^{2}$
with positive diagonal elements $\sigma_{i}^{2}$, whose positive
square root is in turn denoted by $\sigma_{i}$. We also introduce
the column vector $\sigma=\left(\sigma_{1},\ldots,\sigma_{n}\right)'$
such that $\Sigma=\text{diag}\left(\sigma_{1},\ldots,\sigma_{n}\right)$.
While the components of $\varepsilon_{t}$ at one point in time are
also cross-sectionally independent, they may follow different univariate
distributions. \textcolor{red}{}We assume that the stability condition
\begin{equation}
\det\left(a(z)\right)\neq0,\ \left|z\right|\leq1,\label{eq:stability}
\end{equation}
holds, and that there are no determinantal zeros of $b(z)$ on the
unit circle\footnote{Determinantal zeros of $b(z)$ correspond to unit canonical correlations
between the future $\left(y_{t},y_{t+1},\ldots\right)$ and the past
$\left(y_{t-1},y_{t-2},\ldots\right)$ of a stationary stochastic
process \citep{hannanposkitt1988}. Therefore, it seems reasonable
to exclude this case from analysis.}, i.e.
\begin{equation}
\det\left(b(z)\right)\neq0,\ \left|z\right|=1\label{eq:no_unitcircle_zeros}
\end{equation}
hold, and that $B$ is invertible. We will discuss identifiability
of $\left(b(z),B,\Sigma\right)$ in detail in the next subsections
and in Section \ref{sec:identification_scheme}.

The (strictly) stationary solution $\left(y_{t}\right)_{t\in\mathbb{Z}}$
of the system \eqref{eq:system} is called an ARMA process.

Furthermore, we assume that the polynomial matrices $a(z)$ and $b(z)$
are left-coprime\footnote{Two matrix polynomials are called left-coprime if $\left(a(z),b(z)\right)$
is of full row rank for all $z\in\mathbb{C}$. For equivalent definitions
see \citet[Lemma 2.2.1 , page 40]{HannanDeistler12}.}, that $a_{p}$ and $b_{q}$ are non-zero, and that $\left(a_{p},b_{q}\right)$
is of full rank\footnote{The stability, coprimeness, and full-rank assumptions on the parameters
in $a(z)$ and $b(z)$ could be relaxed. The full rank assumption
on $\left(a_{p},b_{q}\right)$ is over-identifying in the sense that
some rational transfer function cannot be parameterised by any VARMA(p,q)
system which satisfies this assumption, see \citet{Hannan71} or \citet[ Chapter 2.7, page 77]{HannanDeistler12}.
To solve this problem, one could consider the parameter space where
the column degrees of $\left(a(z),b(z)\right)$ are fixed to be $\left(p_{1},\ldots,p_{n},q_{1},\ldots,q_{n}\right)$
as in \citet{deistler83} or \citet[Chapter 2.7]{HannanDeistler12}.
Be that as it may, we impose slightly stronger assumptions to strike
a balance between notational complexity and generality, and to focus
on the essential part of this contribution. Using non-Gaussianity
to reduce the equivalence class of stable SVARMA models which generate
the same second moments.}. We do not require that $b_{0}$ be equal to the identity matrix
or non-singular\footnote{While requiring that the $b_{0}$ matrix be non-singular and subsequently
normalising it to the identity matrix is sometimes useful for making
connections to better known time series models and for simplifying
arguments in proofs, such an assumption is unnatural in the context
of the WHF.}. Zeros of $b(z)$ at zero are used for modelling information flows
where that information arrives later to an outside observer than it
does for the economic agents.
\begin{rem}
An assumption similar to the full rank assumption on $\left(a_{p},b_{q}\right)$
seems to be missing in \citet{GourierouxMR_svarma19}. While assuming
coprimeness reduces the equivalence class of SVARMA models $\left(a(z),b(z)B\right)$
that generate the same transfer function $k(z)=a(z)^{-1}b(z)B$, it
is not sufficient to guarantee that the equivalence class is a singleton.
For example, if $u_{1}\left(a_{p},b_{q}\right)=0$ and $u_{1}\neq0$,
then for $\tilde{u}(z)=I_{n}+u_{1}z$ the pair $\left(\tilde{u}(z)a(z),\tilde{u}(z)b(z)B\right)$
is another realisation of the transfer function $k(z)=\left(\tilde{u}(z)a(z)\right)^{-1}\left(\tilde{u}(z)b(z)B\right)$
which satisfies all requirements on the parameter space. Note that
these authors require that, in my notation, $b_{0}=I_{n}$ and therefore
do not allow for delays regarding arrival of information to outside
observers.
\end{rem}

\subsection{Parametrisation using the Wiener-Hopf Factorisation}

The following parametrisation of the MA polynomial matrix $b(z)$
is useful for gaining structural insights into the behaviour of the
system and for deriving asymptotic properties and analytic expressions
for the score, and the information matrix. Every $b(z)=b_{0}+b_{1}z+\cdots+b_{q}z^{q}$
without zeros on the unit circle can be represented as a product of
a backward, a shift and a forward part such that $b(z)=p(z)s(z)f(z)$
where the polynomial matrix $p(z)=p_{0}+p_{1}z+\cdots+p_{q_{p}}z^{q_{p}}$
has no zeros inside or on the unit circle, $s(z)$ is a diagonal matrix
with diagonal entries of the form $z^{\kappa_{i}}$, where $\kappa_{1}\geq\cdots\geq\kappa_{n}$
holds for the so-called partial indices $\kappa_{i}\in\mathbb{Z}$,
and $f(z)=f_{0}+f_{1}z^{-1}+\cdots+f_{q_{f}}z^{-q_{f}}$ has no zeros
or poles outside the unit circle - in particular, it has no zeros
or poles at infinity\footnote{In the univariate case, a polynomial of degree $d$ has $d$ poles
at infinity. }.

Here we provide simple definitions of finite and infinite zeros and
poles of a square matrix $R(z)$ whose elements are rational functions
and whose determinant is not identically zero. While these definitions
suffice for understanding the factorisation mentioned above, we will
discuss different definitions of finite and infinite zeros and poles
(in a more general setting) in the Appendix.

A finite pole of $R(z)$ at $z_{0}\in\mathbb{C}$ is defined as a
point for which an element of $R(z)$ has a pole. At points where
$R(z)$ does not have a pole, $R(z)$ has a finite zero at $z_{0}$
if and only if $\det\left(R(z_{0})\right)=0$. More generally, $R(z)$
has a zero at $z_{0}$ if and only if $R(z)^{-1}$ has a pole at $z_{0}$.

Regarding the point at infinity, $R(z)$ has a pole at infinity if
any element is unbounded when $\left|z\right|\rightarrow\infty$,
or equivalently, if $R\left(\frac{1}{z}\right)$ has a pole at zero.
If there is no pole at infinity, it has a zero at infinity if and
only if the determinant of $R\left(\frac{1}{z}\right)$ is zero when
evaluated at zero. Otherwise, $R(z)$ has a zero at infinity if and
only if any element of $\left(R\left(\frac{1}{z}\right)\right)^{-1}$
has a pole at zero.

Notice that $f(z)$ having no pole at infinity implies that $\left.f\left(\frac{1}{z}\right)\right|_{z=0}$
is finite (or equivalently that $\lim_{\left|z\right|\rightarrow\infty}f(z)$
is finite)\footnote{In system theory, a rational matrix function satisfying $\lim_{\left|z\right|\rightarrow\infty}R(z)<\infty$
or $\lim_{\left|z\right|\rightarrow\infty}R(z)=0$ is called proper
or strictly proper. The latter is often used for finding a system
realisation of the transfer function since it is easy to build a state
space system $\left(A,B,C\right)$ from a strictly proper $R(z)=C\left(zI-A\right)^{-1}B$
and subsequently obtain a proper one as $\left(C\left(z-A\right)^{-1}BD^{-1}+I\right)D$.}. Moreover, $f(z)$ not having infinite zeros implies that $\left.f\left(\frac{1}{z}\right)\right|_{z=0}=f_{0}$
is of full rank.

\subsubsection{Existence of the WHF}

The factorisation of $b(z)$ into $\left(p(z),s(z),f(z)\right)$ is
known as the Wiener-Hopf factorisation (WHF) \citep{gohberg_feldman81,clanceygohberg81,gohkaaspit03_summerschool},
see also \citet{onatski06,AlSadoon18_ET_lrem,alsadoon2019identification}\textcolor{red}{}
for its use in rational expectations models. Every rational matrix
function without determinantal zeros on the unit circle admits a WHF
\citep[Chapter I]{clanceygohberg81}. It is not necessary (or even
unnatural) to assume that $b_{0}=I_{n}$ or non-singular when using
the WHF to parametrise the MA polynomial matrix.

Explicit construction of the (left-) WHF\footnote{A similar construction can be found in the Supplementary Appendix
of \citet{AlSadoon18_ET_lrem}.} of $b(z)$ using the Smith form \citep[page 313ff.]{GohbergLancasterRodman09}
provides also insights into the relation of the WHF to other (better
known) factorisations of polynomial matrices. We start from the matrix
polynomial $b(z)=b_{0}+b_{1}z+\cdots+b_{q}z^{q}$ and obtain 
\[
b(z)=\underbrace{\left[u(z)\Lambda_{p}(z)\right]}_{=\tilde{\tilde{p}}(z)}\underbrace{\left[\Lambda_{f}(z)v(z)\right]}_{=\tilde{\tilde{\tilde{f}}}(z)}=\underbrace{\left[\tilde{\tilde{p}}(z)w(z)^{-1}\right]}_{=\tilde{p}(z)}\underbrace{\left[w(z)\tilde{\tilde{\tilde{f}}}(z)\right]}_{=\tilde{\tilde{f}}(z)}
\]
where $\Lambda_{p}(z)$ has only zeros outside the unit circle, and
$\Lambda_{f}(z)$ has only zeros inside the unit circle, and $w(z)$
is a unimodular matrix which row-reduces\footnote{See the Appendix for the definition of row-reduced polynomial matrices.
E.g. a polynomial matrix $M(z)$ whose coefficient matrix pertaining
to the highest power of $z$ is non-singular is row- and column-reduced.} $\tilde{\tilde{\tilde{f}}}(z)$, see \citet[Theorem 2.5.7, page 28]{wolovich74},
\citet[page 386]{Kailath1980}, and \citet{basilio02}. Subsequently,
we permute the rows of $\tilde{\tilde{f}}(z)$ such that for the row
degrees $\kappa_{i}$ the inequalities $\kappa_{1}\geq\cdots\geq\kappa_{n}$
hold and we extract the highest degree of each row to obtain the partial
indices
\[
b(z)=\underbrace{\left[\tilde{p}(z)P'\right]}_{=p(z)}\underbrace{\left[P\tilde{\tilde{f}}(z)\right]}_{=\tilde{f}(z)}=p(z)\underbrace{\text{diag}\left(z^{\kappa_{1}},\ldots,z^{\kappa_{n}}\right)}_{=s(z)}\underbrace{\left[\text{diag}\left(z^{-\kappa_{1}},\ldots,z^{-\kappa_{n}}\right)\tilde{f}(z)\right]}_{=f(z)}
\]
Note that $f(z)$ does not have poles at infinity since its degree
is zero and that it does not have zeros at infinity because $f\left(\frac{1}{z}\right)$
evaluated at $z=0$ is by construction of full rank. In particular,
$f_{0}$ as well as $p_{0}$ are non-singular. However, this is not
necessarily the case for $b_{0}$.

\subsubsection{Genericity of WHF}

For an open and dense set in the set of all feasible MA polynomial
matrices $b(z)$, the partial indices are such that the difference
between the largest one $\kappa_{1}$ and the smallest one $\kappa_{n}$
is at most one. Moreover, partial indices of this form are stable
in the sense that they do not change under small perturbations. This
has been proved in \citet[page 260ff.]{GohbergKrein60}. The following
intuitive description of this result is based on \citet[Section 1.5]{gohkaaspit03_summerschool}.

We start with an example illustrating the mechanics of small perturbations.
The partial indices of $b_{\varepsilon}(z)=\left(\begin{smallmatrix}z^{2} & 0\\
\varepsilon\cdot z & 1
\end{smallmatrix}\right)$ for $\varepsilon=0$ are obviously $\left(\kappa_{1}^{(\varepsilon)},\kappa_{2}^{(\varepsilon)}\right)=\left(2,0\right)$.
For $\varepsilon\neq0$, its WHF is $b_{\varepsilon}(z)=\left(\begin{smallmatrix}z & 1\\
\varepsilon & 0
\end{smallmatrix}\right)\left(\begin{smallmatrix}z & 0\\
0 & z
\end{smallmatrix}\right)\left(\begin{smallmatrix}1 & \frac{z^{-1}}{\varepsilon}\\
0 & \frac{1}{\varepsilon}
\end{smallmatrix}\right)$ such that $\left(\kappa_{1}^{(\varepsilon)},\kappa_{2}^{(\varepsilon)}\right)=\left(1,1\right)$.
We say that partial index $\kappa$ majorities partial index $\mu$,
if $\mu$ can be obtained from $\kappa$ by finitely many elementary
changes of the form $\mu_{j}=\kappa_{j}-1$ and $\mu_{k}=\kappa_{k}+1$
for $j<k$ (including the identity transformation as well as a possible
reordering) and define in this way a partial order $\kappa\succ\mu$
on the set of all partial indices\footnote{Note that for $\mu$ obtained from $\kappa$ by elementary operations
it holds that $\sum_{i=1}^{n}\kappa_{i}=\sum_{i=1}^{n}\mu_{i}$ .}. According to Theorem 1.20 in \citep[Section 1.5]{gohkaaspit03_summerschool},
for a given MA matrix polynomial $b(z)$ with partial indices $\kappa$,
every neighbourhood contains an MA polynomial matrix with partial
indices $\mu$ which are majored by $\kappa$. According to Theorem
1.21 in \citep[Section 1.5]{gohkaaspit03_summerschool}, for a given
MA polynomial matrix with arbitrary partial indices $\left(\kappa_{1},\ldots,\kappa_{n}\right)$,
there is a neighbourhood in which for all other MA matrix polynomials
it holds that their partial indices are majored by $\kappa.$

Therefore, the set of partial indices which is stable under small
perturbations is the one which is minimal with respect to the partial
order $\succ$. For each such partial index $\mu,$ it holds that
$\mu_{1}\leq\mu_{n}+1$ such that $\mu=\left(\mu_{1},\ldots,\mu_{n}\right)=\left(\kappa+1,\ldots,\kappa+1,\kappa,\ldots,\kappa\right)$
may be described by two integers $\left(\kappa,k\right)$, $0\leq\kappa\leq q$
and $k\in\left\{ 0,\ldots,n-1\right\} $, which satisfy $n\cdot\kappa+k=\sum_{i=1}^{n}\mu_{i}$
such that $\left(\mu_{1},\ldots,\mu_{n}\right)=\left(\kappa+1,\ldots,\kappa+1,\kappa,\ldots,\kappa\right)$,
i.e. the first $k$ partial indices are equal to $\kappa+1$ and the
last $(n-k)$ ones are equal to $\kappa$.

The above results are summarised in
\begin{prop}[\citet{GohbergKrein60}]
\label{thm:whf_known}Every matrix polynomial $b(z)=b_{0}+b_{1}z+\cdots+b_{q}z^{q}$
in an open and dense subset of the open set $\bigcup_{\left|z_{0}\right|=1}\left\{ \left(b_{0},b_{1},\ldots,b_{q}\right)\in\mathbb{R}^{n^{2}\left(q+1\right)}\,|\,\det\left(b\left(z_{0}\right)\right)\neq0\right\} \subseteq\mathbb{R}^{n^{2}(q+1)}$
has a WHF $b(z)=p(z)s(z)f(z)$ where $p(z)$ has no zeros or poles
inside or on the unit circle, $s(z)=\text{diag}\left(z^{\kappa+1},\ldots,z^{\kappa+1},z^{\kappa},\ldots z^{\kappa}\right)$
with $\left(\kappa,k\right)$ such that $0\leq\kappa\leq q-1$ and
$k\in\left\{ 0,\ldots,n-1\right\} $ or $\left(\kappa,k\right)=\left(q,0\right)$
and $f(z)$ has no zeros or poles outside or on the unit circle.
\end{prop}
\begin{rem}
One of the shortcomings of using an approach based on Blaschke matrices
like \citet{GourierouxMR_svarma19} or \citet{velasco2020identification}
is the specification of the parameter space in terms of zero locations
of all MA zeros. Even for a univariate polynomial, it is in general
not possible to find continuous functions (with respect to the parameter
vector) for its zeros \citet[page 373]{HinrichsenPritchard00}. In
the multivariate case, this issue is even more complex since there
are two cases to be distinguished when the determinant of a polynomial
matrix has a zero of multiplicity two at, say, $z=\alpha$. Indeed,
the Smith-form of such a bivariate polynomial matrix may be of the
form $\left(\begin{smallmatrix}1 & 0\\
0 & (z-\alpha)^{2}
\end{smallmatrix}\right)$ or $\left(\begin{smallmatrix}z-\alpha & 0\\
0 & z-\alpha
\end{smallmatrix}\right)$. Specification of well-behaved parameter spaces is a complex task,
see \citet[Chapter 2.5 and 2.6]{HannanDeistler12}, and potential
pitfalls are described in \citet[page 128, Remark 4]{HannanDeistler12}.
\end{rem}
\begin{rem}
While in the univariate treatment of ARMA models in \citet[Assumption A(p,q), page 562]{VelascoLobato18},
MA polynomials which have both zeros at $z=\alpha$ and $z=\frac{1}{\bar{\alpha}}$
are excluded, a similar assumption seems to be missing in \citet{velasco2020identification},
see also \citet{BaggioFerrante16} who exclude this case when they
analyse the rational spectral factorisation problem. \citet{GourierouxMR_svarma19}
do not address properties of their parameterisation (other than the
incorrect statement that the parameter space for VARMA models is compact).
\end{rem}

\subsubsection{(Non-) Uniqueness of the WHF}

While the WHF is not unique, we will now show how to construct a unique
representative from the class of all WHF for a given MA polynomial
$b(z)$.

In the case $\left(\kappa,0\right)$, the WHF is essentially unique
in the sense that the equivalence class of WHFs for $b(z)$ is parametrised
by the set of non-singular (constant) matrices of dimension $\left(n\times n\right)$.
In particular, requiring that $p(0)=I_{n}$ results in a unique WHF
of $b(z)$.

In the case $\left(\kappa,k\right)$, $k\neq0$, we first show that
the the first $k$ columns of $p(z)$ have degree smaller than or
equal to $q-\kappa-1$ and the last $n-k$ columns of $p(z)$ have
degree smaller than or equal to $q-\kappa$ by using the predictable
degree property \citep[Theorem 6.3-13, page 387]{Kailath1980}:
\begin{lem}
Let $M(z)$ be an $\left(n\times n\right)$-dimensional polynomial
matrix whose determinant is not identically zero. For every polynomial
row vector $w(z)$, let $v(z)=w(z)M(z)$. $M(z)$ is row-reduced if
and only if $\deg\left(v(z)\right)=\max_{j}\left\{ \deg\left(w_{j}(z)\right)+\deg\left(M_{[j,\bullet]}\right)\right\} $.

\end{lem}
It follows from the construction of the WHF using the Smith form that
$g(z):=s(z)f(z)$ is row-reduced. The row degrees of $g(z)$ are $\left(\kappa+1,\ldots,\kappa+1,\kappa,\ldots,\kappa\right)$
and since the degrees of the rows of $b(z)$ are smaller than or equal
to $q$, it follows for each row of $p(z)$ that the degrees of the
first $k$ elements are bounded by $q-\kappa-1$ and the degrees of
the last $n-k$ elements are bounded by $q-\kappa$. 

Next, we use the fact \citep[Theorem I.1.2, page 11]{clanceygohberg81}
that the equivalence class of WHFs for $b(z)$ is parametrised by
the block upper triangular unimodular matrices for which $u_{[k+1:n,1:k]}(z)=0$,
the diagonal blocks are constant, and the degree of $u_{[1:k,k+1:n]}(z)$
is at most one. Note that this unimodular transformation neither change
the row degrees of $f(z)$ nor the column degrees of $p(z)$. We choose
a canonical representative among all pairs\footnote{The partial indices (and therefore $s(z)$) are unique.}
$\left(p(z),f(z)\right)$ for given partial indices $\left(\kappa,k\right)$
by requiring that the (non-singular) zero-lag coefficient matrix $p_{0}$
be equal to $\left(\begin{smallmatrix}I_{k} & 0_{k\times(n-k)}\\
* & I_{n-k}
\end{smallmatrix}\right)$, where the asterisk denotes unrestricted elements, and that\footnote{The matrix $u_{0}$ in $u(z)=u_{0}+\begin{pmatrix}0 & \tilde{u}_{1}\\
0 & 0
\end{pmatrix}z$ is determined as follows. Let us partition the matrix $p_{0}=\begin{pmatrix}p_{0,11} & p_{0,12}\\
p_{0,21} & p_{0,22}
\end{pmatrix}$ and assume without loss of generality that $p_{0,11}$ is invertible.
Then, right-multiplying $p(z)$ with $u_{0}=\left(\begin{smallmatrix}p_{0,11}^{-1} & 0\\
0 & I_{n-k}
\end{smallmatrix}\right)\left(\begin{smallmatrix}I_{k} & -p_{0,12}\\
0 & I_{n-k}
\end{smallmatrix}\right)\left(\begin{smallmatrix}I_{k} & 0\\
0 & \left(p_{0,22}-p_{0,21}p_{0,11}^{-1}p_{0,12}\right)^{-1}
\end{smallmatrix}\right)$ leads to
\begin{align*}
\begin{pmatrix}p_{0,11} & p_{0,12}\\
p_{0,21} & p_{0,22}
\end{pmatrix}u_{0} & =\begin{pmatrix}I_{k} & p_{0,12}\\
p_{0,21}p_{0,11}^{-1} & p_{0,22}
\end{pmatrix}\begin{pmatrix}I_{k} & -p_{0,12}\\
0 & I_{n-k}
\end{pmatrix}\begin{pmatrix}I_{k} & 0\\
0 & \left(p_{0,22}-p_{0,21}p_{0,11}^{-1}p_{0,12}\right)^{-1}
\end{pmatrix}\\
 & =\begin{pmatrix}I_{k} & 0\\
p_{0,21}p_{0,11}^{-1} & p_{0,22}-p_{0,21}p_{0,11}^{-1}p_{0,12}
\end{pmatrix}\begin{pmatrix}I_{k} & 0\\
0 & \left(p_{0,22}-p_{0,21}p_{0,11}^{-1}p_{0,12}\right)^{-1}
\end{pmatrix}=\begin{pmatrix}I_{k} & 0\\
p_{0,21}p_{0,11}^{-1} & I_{n-k}
\end{pmatrix}.
\end{align*}
In order to fix $\tilde{u}_{1}$, we require that $p_{1,12}=0$.} $p_{1,[1:k,k+1:n]}=0$.

We may thus construct a canonical representative of a simple form
by restricting certain parameters to zero and one and summarise this
in
\begin{thm}
\label{thm:whf}In the case $k=0,$ a unique representative among
all WHF of $b(z)$ can be chosen by requiring that $p_{0}=I_{n}$
where $p(z)=I_{n}+p_{1}z+\cdots+p_{q-\kappa}z^{q-\kappa}$ and $f(z)=f_{0}+f_{1}z^{-1}+\cdots+f_{\kappa}z^{-\kappa}$.
In the case $k\neq0$, we have that $\deg\left(p_{[\bullet,1:k]}(z)\right)=q-\kappa-1$,
$\deg\left(p_{[\bullet,k+1:n]}(z)\right)=q-\kappa$, and that $f(z)=f_{0}+f_{1}z^{-1}+\cdots+f_{\kappa+1}z^{-\kappa-1}$
where and $f_{\kappa+1,[k+1:n,\bullet]}=0$. In this case, a unique
representative can be chosen by requiring that $p_{0}=\left(\begin{smallmatrix}I_{k} & 0\\
p_{0,21} & I_{n-k}
\end{smallmatrix}\right)$ and $p_{1,12}=0_{k\times(n-k)}$.
\end{thm}
\begin{rem}
There are $n\cdot\kappa+k$ zeros inside the unit circle and $\deg\left(\det\left(b(z)\right)\right)-\left(n\cdot\kappa+k\right)$
zeros outside the unit circle.
\end{rem}
\begin{rem}
The number of free parameters only depends on $\left(p,q\right)$
and is independent of the partial indices $\left(\kappa,k\right)$.
\end{rem}
From now on, we assume that this canonical representative among all
WHF factorisations of $b(z)$ has been chosen. Next, we discuss two
different ways for normalising $b(z)=p(z)s(z)f(z)=b_{0}+\cdots+b_{q}z^{q}$.
In the first ``natural'' normalisation, we require that $f_{0}=I_{n}$,
in the second one we require that $\left(\begin{smallmatrix}f_{\kappa+1,[1:k,\bullet]}\\
f_{\kappa,[k+1:n,\bullet]}
\end{smallmatrix}\right)$ is firstly of full rank and secondly equal to the inverse of the
$p_{0}$. The former normalisation is natural for modelling MA matrix
polynomials with zeros inside the unit circle (including at zero).
The latter one is more restrictive since it excludes zeros at zero
and implies that $b_{0}=I_{n}$. Note that the assumption $b_{0}=I_{n}$
is unnecessary for guaranteeing the existence of the WHF and that
the matrix $B$ in the right-hand-side of equation \eqref{sec:Model}
corresponds to the zero-lag coefficient matrix in this case.
\begin{cor}
\label{cor:unique_whf}The unique representative of the WHF of the
matrix polynomial $b(z)$ of Theorem \ref{thm:whf} may be further
factored such that $b(z)=\left(p(z)s(z)f^{(1)}(z)\right)f_{0}$ where
$f^{(1)}(z)=f(z)f_{0}^{-1}$ such that $\left.f^{(1)}\left(\frac{1}{z}\right)\right|_{z=0}=I_{n}$.
Under the additional assumption that $b_{0}$ be non-singular, we
may also factor $b(z)$ such that $b(z)=\left(p(z)s(z)f^{(2)}(z)\right)b_{0}$
where $\left(\begin{smallmatrix}f_{\kappa+1,[1:k,\bullet]}^{(2)}\\
f_{\kappa,[k+1:n,\bullet]}^{(2)}
\end{smallmatrix}\right)=\left(\begin{smallmatrix}I_{k} & 0\\
-p_{0,21} & I_{n-k}
\end{smallmatrix}\right)$ holds in the case $k\neq0$ and where $f_{\kappa}^{(2)}=I_{n}$ in
the case $k=0$. The other coefficient matrices in $f^{(2)}(z)$ remain
unrestricted.
\end{cor}
\begin{example}
For illustrating the non-uniqueness as well as different normalisation
choices of the WHF consider a bivariate MA polynomial matrix $b(z)$
of degree $q=3$ and with partial index $\left(\kappa,k\right)=\left(1,1\right)$
\begin{gather*}
b(z)=p(z)s(z)f(z)=\\
=\left[\begin{pmatrix}1 & \frac{1}{4}\\
\frac{1}{3} & \frac{1}{2}
\end{pmatrix}+\begin{pmatrix}\frac{1}{2} & \frac{1}{3}\\
\frac{1}{4} & \frac{1}{3}
\end{pmatrix}z+\begin{pmatrix}0 & \frac{1}{3}\\
0 & \frac{1}{2}
\end{pmatrix}z^{2}\right]\begin{pmatrix}z^{2} & 0\\
0 & z
\end{pmatrix}\left[\begin{pmatrix}1 & 2\\
3 & 4
\end{pmatrix}+\begin{pmatrix}\frac{1}{2} & \frac{1}{5}\\
\frac{1}{4} & \frac{1}{3}
\end{pmatrix}z^{-1}+\begin{pmatrix}\frac{1}{6} & \frac{1}{7}\\
0 & 0
\end{pmatrix}z^{-2}\right].
\end{gather*}
The unimodular matrix $u(z)$ parametrising the equivalence class
of all WHFs for a given MA polynomial matrix is chosen to be $u(z)=\left(\begin{smallmatrix}1 & -\frac{3}{5}-\frac{1}{2}z\\
0 & \frac{12}{5}
\end{smallmatrix}\right)$ such that we obtain the unique representative among the equivalence
class of all WHF for $b(z)$, i.e.
\begin{gather*}
b(z)=\tilde{p}(z)s(z)\tilde{f}(z)=\\
=\left[\begin{pmatrix}1 & 0\\
\frac{1}{3} & 1
\end{pmatrix}+\begin{pmatrix}\frac{1}{2} & 0\\
\frac{1}{4} & \frac{29}{60}
\end{pmatrix}z+\begin{pmatrix}0 & \frac{11}{20}\\
0 & \frac{43}{40}
\end{pmatrix}z^{2}\right]\begin{pmatrix}z^{2} & 0\\
0 & z
\end{pmatrix}\left[\begin{pmatrix}\frac{13}{8} & \frac{17}{6}\\
\frac{5}{4} & \frac{5}{3}
\end{pmatrix}+\begin{pmatrix}\frac{125}{96} & \frac{457}{360}\\
\frac{5}{48} & \frac{5}{36}
\end{pmatrix}z^{-1}+\begin{pmatrix}\frac{19}{48} & \frac{1}{3}\\
0 & 0
\end{pmatrix}z^{-2}\right],
\end{gather*}
where $\tilde{p}(z)s(z)\tilde{f}(z)=\left\{ p(z)u(z)\right\} s(z)\left\{ \left[s^{-1}(z)u^{-1}(z)s(z)\right]f(z)\right\} $.

Regarding normalisation, note that $b_{0}$ is equal to $\tilde{p}_{0}\cdot\left(\begin{smallmatrix}\tilde{f}_{2,[1,\bullet]}\\
\tilde{f}_{1,[2,\bullet]}
\end{smallmatrix}\right)=\left(\begin{smallmatrix}1 & 0\\
\frac{1}{3} & 1
\end{smallmatrix}\right)\left(\begin{smallmatrix}\frac{19}{48} & \frac{1}{3}\\
\frac{5}{48} & \frac{5}{36}
\end{smallmatrix}\right)$. Since $b_{0}$ is non-singular, we could further transform $\tilde{f}(z)$
by post-multiplication with a non-singular matrix to $\tilde{f}^{(2)}(z)$
such that $\left.\left(\tilde{p}(z)s(z)\tilde{f}^{(2)}(z)\right)\right|_{z=0}=I_{2}$
and write $b(z)=\left(\tilde{p}(z)s(z)\tilde{f}^{(2)}(z)\right)b_{0}$.
Likewise, and without imposing any assumption on $b_{0}$, we may
post-multiply $\tilde{f}(z)$ with $\left(\tilde{f}_{0}\right)^{-1}$
and obtain $b(z)=\left(\tilde{p}(z)s(z)\tilde{f}^{(1)}(z)\right)\tilde{f}_{0}^{(1)}$.
\end{example}

Next, we point out differences and similarities of the WHF to the
factorisation of the AR matrix polynomial of the form $\tilde{a}(z)=\Pi(z)\Phi\left(\frac{1}{z}\right)=\left(I-\Pi_{1}z-\cdots-\Pi_{r}z^{r}\right)\left(I-\Phi_{1}z^{-1}-\cdots-\Phi_{s}z^{-s}\right)$
in \citet{LanneSaikkonen13} in order to assess the statement in \citep[pages 124 and 125]{goujas17_noncausal_semiparam}
that the factorisation in \citet{LanneSaikkonen13} is ``very restrictive''.
Both $\det\left(\Pi(z)\right)$ and $\det\left(\Phi(z)\right)$ have
no zeros inside or on the unit circle. Thus, $\Pi(z)$ and $\Phi\left(\frac{1}{z}\right)$
correspond to our $p(z)$ and $f(z)$ where $f(z)$ is normalised
``in the natural way'' and the normalisation factor is incorporated
in the error covariance matrix in \citet{LanneSaikkonen13}. This
factorisation is less general than the (generic) WHF to the extent
that all partial indices must be equal. Moreover, identification has
to rely on cross-sectionally dependent inputs $\varepsilon_{t}$,
see \citet[Proposition 1, page 457]{LanneSaikkonen13} based on \citet[Theorem 7]{chanho04},
since their $\tilde{a}(z)$ is not a matrix polynomial but rather
a matrix Laurent polynomial featuring negative and positive powers
of $z$.

\section{Identifiability Analysis}

We follow \citet{Rothenberg71} and \citet{DeistlerSeifert78} to
define identifiability of parametric models. The external characteristic
of the stationary solution $\left(y_{t}\right)_{t\in\mathbb{Z}}$
of \eqref{eq:system} is the probability distribution function (or
a subset of corresponding moments). A particular system \eqref{eq:system}
is described by the parameters of \eqref{eq:system} which satisfy
assumptions \eqref{eq:stability} and \eqref{eq:no_unitcircle_zeros}
as well as the coprimeness assumption, the full rank assumption and
the assumptions on $B$ and $\Sigma$. The model is then characterised
by the set of all a priori possible systems which we will call internal
characteristics. Two systems of the form \eqref{eq:system} are called
observationally equivalent if they imply the same external characteristics
of $\left(y_{t}\right)_{t\in\mathbb{Z}}$. A system is identifiable
if there is no other observationally equivalent system. The identifiability
problem is concerned with the existence of an injective function from
the internal characteristics to the external characteristics\footnote{The inverse of this function, i.e. from the external to the internal
characteristics, is called the identifying function.}, see \citet{DeistlerSeifert78} for a more detailed discussion.

\subsection{Non-Identifiability from Second Moments and Methods based on Blaschke
Matrices}

The classical (non-)identifiability issues where the external characteristics
are described by the second moments of $\left(y_{t}\right)_{t\in\mathbb{Z}}$
are best understood in terms of the spectral density of the stationary
solution of \eqref{eq:system}. The spectral density, i.e. the Fourier
transform of the autocovariance function $\gamma(s)=\mathbb{E}\left(y_{t}y_{t-s}'\right),\ s\in\mathbb{Z},$
of $\left(y_{t}\right)_{t\in\mathbb{Z}}$ , is 
\[
f(z)=\frac{1}{2\pi}\sum_{\alpha=-\infty}^{\infty}\gamma(\alpha)z^{\alpha}=a(z)^{-1}b(z)B\Sigma^{2}B'b'\left(\frac{1}{z}\right)a'\left(\frac{1}{z}\right)^{-1},
\]
evaluated at $z=e^{-i\lambda}$, $\lambda\in\left[-\pi,\pi\right]$.

Starting identifiability analysis from this rational spectral density
without zeros on the unit circle, it is well known \citep[Theorem 10.1, page 47]{Rozanov67},
\citep[Theorem II.10' page 66 and Theorem III.1 on page 129]{Hannan70},
\citep{BaggioFerrante16}, that there exists a canonical rational
spectral factor $l(z)$ without zeros or poles on or inside the unit
circle such that $f(z)=l(z)l'\left(\frac{1}{z}\right)$. This canonical
spectral factor is unique up to orthogonal post-multiplication. In
order to focus on the non-uniqueness implied by different pole and
zero locations, we will for now abstract from the ``static'' non-uniqueness
of spectral factors implied by orthogonal post-multiplication on $l(z)$
by requiring that the coefficient pertaining to power zero of $z$
in the respective spectral factor is lower-triangular with positive
diagonal elements.

When allowing for spectral factors with unrestricted zero and pole
location, there generally exists an infinite number of rational all-pass
filters $V(z)$, which satisfy $V(z)V'\left(\frac{1}{z}\right)=I_{n}$
\citep[page 207]{alpgohberg88}, such that $f(z)=\left[l(z)V(z)\right]V'\left(\frac{1}{z}\right)l'\left(\frac{1}{z}\right)=\tilde{l}(z)\tilde{l}'\left(\frac{1}{z}\right)$
holds. Requiring that the spectral factors with arbitrary pole and
zero location be minimal\footnote{A spectral factor is minimal if the number of its finite and infinite
poles (including multiplicities) is one half of the number of finite
and infinite poles (including multiplicities) of the spectral density,
see the Appendix for the definition of zeros and poles including their
multiplicities and structure using the Smith-McMillan form. This excludes,
e.g., spectral factors that are obtained by post-multiplying the canonical
spectral factor by all-pass filters which do not cancel any zero or
pole of $l(z)$ and which correspond to what \citet{LippiReichlin94}
call ``non-basic representations''.}, \citet{BaggioFerrante19_parametrization_phasefunction} have recently
shown that the finite set of all minimal spectral factors $\tilde{l}(z)$
of $f(z)$ can be obtained by right-multiplying the divisors\footnote{The rational matrices $T_{l}(z)$ and $T_{r}(z)$ are respectively
left all-pass divisor and right all-pass divisor of the rational all-pass
filter $T(z)$ if $T(z)=T_{l}(z)T_{r}(z)$ holds and there are no
(finite or infinite) pole or zero cancellations between $T_{l}(z)$
and $T_{r}(z)$.} of a particular rational all-pass filter $T(z)$ on the canonical
spectral factor $l(z)$. We may obtain $T(z)=l(z)^{-1}j(z)$ from
the canonical spectral factor $l(z)$ (without zeros and poles inside
or on the unit circle) and another ``extremal'' spectral factor
$j(z)$ which has no zeros and poles outside or on the unit circle.
Since $l(z)l'\left(\frac{1}{z}\right)=j(z)j'\left(\frac{1}{z}\right)$,
it is clear that $T(z)$ is indeed all-pass. Moreover, the all-pass
filter $T(z)$ may be represented as the product of orthogonal matrices
and so-called Blaschke matrices of the form $\left(\begin{smallmatrix}I_{r} & 0_{r\times(n-r)}\\
0_{(n-r)\times r} & I_{n-r}\frac{1-\bar{\alpha}z}{z-\alpha}
\end{smallmatrix}\right)$, see \citet[page 65]{Hannan70}, \citet[Theorem 1, page 311]{LippiReichlin94},
or \citet[Theorem 3.12, page 208]{alpgohberg88}, which immediately
provides the (finite number of) all-pass divisors of $T(z)$ which
in turn generate a finite number of minimal spectral factors with
different zero and pole locations. Of course, one needs to ensure
that the obtained spectral factors have real-valued coefficients.
Thus, complex-conjugated zeros of $l(z)$ have to be mirrored jointly
into the unit circle. Note that constructing an all-pass filter with
real-valued coefficients is non-trivial \citep{scherrer_funovits2020allpass}
and that naive approaches like in \citet{GourierouxMR_svarma19} lead
to all-pass filters and spectral factors with complex-valued coefficients,
see \citet{funovits2020gmr_comment} for a detailed analysis.

Assuming that one knows the true canonical spectral factor $l(z)$
on the population level, it would be possible to construct $2^{n_{r}+n_{q}}-1$,
where $n_{r}$ is the number of real-valued zeros of $l(z)$ and $n_{q}$
is the number of complex-conjugated pairs of zeros, different spectral
factors which generate the same spectral density and which have real-valued
coefficients. When working with finite realisations of the data generating
process, however, the fact that one might end up with complex-conjugated
roots only (e.g. due to estimation uncertainty) even though the true
spectral factor has only real-valued zeros is problematic for estimation
strategies based on mirroring zeros with Blaschke matrices. In this
case, the number of zeros inside the unit circle is necessarily even.
Note that the method based on the WHF parametrisation presented in
this article does not suffer from this shortcoming and searches for
an optimum over the whole parameter space.

\subsection{\label{sec:identification_scheme}Identifiability using Additional
Information}

We will show that under two different sets of assumptions on the joint
distribution of the components of the inputs $\left(\varepsilon_{t}\right)$
to the SVARMA model \eqref{eq:system}, $\left(a(z),p(z),s(z),f(z)\right)$
are unique and $\left(B,\Sigma\right)$ are unique up to signed permutation.

The (non-) uniqueness of the infinite MA representation of multivariate
linear processes driven by non-Gaussian inputs is well understood
in the time series literature and analysed, e.g., in \citet{chanHoTong06,chanho04}\footnote{These articles generalise univariate results in \citet[Theorem 5, page 46]{rosenblatt1985},
\citet{findley86,findley90}, \citet{cheng1990,cheng1992} which in
turn are based on \citet[Theorem 5.6.1, page 178]{Kagan73} and \citet[Theorem 3.1.1, page 89]{Kagan73}}. These insights are applied straight-forwardly in the econometrics
literature in, e.g., \citet{davis10}, \citet{LanneSaikkonen13} and
\citet{GourierouxMR_svarma19} to show that their respective non-causal
and non-invertible models are identified. 

Either of the following two assumptions on the joint distributions
of the components of $\varepsilon_{t}$ is needed for proving the
identifiability result of \citet{chanho04}.
\begin{assumption}[Identically distributed components]
\label{assu:components_identical}The components of $\varepsilon_{t}$
are independent, identically distributed, and non-Gaussian.
\end{assumption}
This assumption does not require the existence of higher order moments.
However, it is considered restrictive. One may allow for cross-sectional
heterogeneity of $\varepsilon_{t}$ if the existence of some higher
order moments is assumed.
\begin{assumption}[Non-zero cumulants]
\label{assu:components_cumu}The components of $\varepsilon_{t}$
are mutually independent (but not necessarily identically distributed).
Each component has a non-zero cumulant of order $r\geq3$ and finite
moments up to order $\tau$, where $\tau$ is an even integer and
larger than or equal to $r.$
\end{assumption}
The requirement that a cumulant of order at least three be non-zero
excludes the Gaussian distribution. Note that in \citet[Theorem 3, page 8]{chanho04},
the authors do not require directly that the components of $\varepsilon_{t}$
be non-Gaussian but only that they be independent and identically
distributed. Their non-Gaussianity follows from the (assumed) non-Gaussianity
of at least one output component.

Finally, let us state the result on identifiability of our model.
\begin{lem}[Chan, Ho (2004)]
Under Assumption \ref{assu:components_identical} or \ref{assu:components_cumu},
the linear process $y_{t}=\sum_{\alpha=-\infty}^{\infty}w_{j}\varepsilon_{t-j}$
where the $\left(n\times n\right)$ -dimensional filter $w(z)$ has
no zeros on the unit circle and satisfies $\sum_{\alpha=-\infty}^{\infty}\left\Vert w_{j}\right\Vert ^{2}<\infty$
is identified up to permutations, scalings, and shifts. Thus, if there
is another representation $y_{t}=\sum_{\alpha=-\infty}^{\infty}\tilde{w}_{j}\tilde{\varepsilon}_{t-j}$
it holds that $\tilde{w}_{\left[\bullet,i\right]j}=\beta_{i}w_{\left[\bullet,\pi(i)\right]j-m_{i}}$
and $\tilde{\varepsilon}_{i,t}=\beta_{i}^{-1}\varepsilon_{\pi(i),t+m_{i}}$
where $\pi$ is a permutation on $\left\{ 1,\ldots,n\right\} $, $\beta_{i}$
are non-zero scalars, and $m_{i}$ are integers.
\end{lem}
\citet{GourierouxMR_svarma19} apply this result straight-forwardly
to the case where $w_{j}=0$ for $j<0$ and invertible $w_{0}$ to
obtain their Proposition 1.
\begin{thm}
Under Assumption \ref{assu:components_identical} or \ref{assu:components_cumu},
and the assumptions outlined below equation \eqref{eq:system}, the
parameters $\left(a(z),p(z)s(z)f(z),B,\Sigma\right)$ in model \eqref{eq:system}
are identifiable up to signed permutations.
\end{thm}
\begin{proof}
We use the fact that in our case $w_{j}=0$ for $j<0$ and that the
partial indices of the WHF are unique. The scalars are fixed by requiring
that $f_{0}=I_{n}$. The shifts are fixed because one cannot post-multiply
columns of $k(z)=a(z)^{-1}p(z)s(z)f(z)B$ with positive powers of
$z$ since this would change the partial indices. Likewise, one cannot
post-multiply columns of $k(z)$ with negative powers of $z$ since
this would contradict the invertibility of $f_{0}$.
\end{proof}

In order to compare this result with the approach in \citet{velasco2020identification},
we define in the following higher order cumulants \citep{leonov1959method}
and higher order spectral densities \citep{Brillinger75,zurb86}.
To fix ideas, we focus on third order cumulants and the third order
spectral density, i.e. the bispectrum. Let $\varphi_{\eta}\left(\alpha_{1},\alpha_{2},\alpha_{3}\right)=\mathbb{E}\left(e^{i\left(\alpha_{1}\eta_{1}+\alpha_{2}\eta_{2}+\alpha_{3}\eta_{3}\right)}\right)$
be the characteristic function of the vector of random variables $\left(\eta_{1},\eta_{2},\eta_{3}\right)$
for which $\mathbb{E}\left(\left|\eta_{j}\right|^{n}\right)<\infty$.
For $\left(\nu_{1},\nu_{2},\nu_{3}\right)\in\mathbb{Z}^{3},\ \nu_{i}\geq0,$
the cumulants are defined as the coefficients $s_{\eta}^{\left(\nu_{1},\nu_{2},\nu_{3}\right)}$
in the Taylor series expansion of $\log\left(\varphi_{\eta}\left(\alpha_{1},\alpha_{2},\alpha_{3}\right)\right)$
around $\alpha=0$, i.e. 
\[
\log\left(\varphi_{\eta}\left(\alpha_{1},\alpha_{2},\alpha_{3}\right)\right)=\sum_{\nu_{1}+\nu_{2}+\nu_{3}\leq n}\frac{i^{\nu_{1}+\nu_{2}+\nu_{3}}}{\nu_{1}!\cdots\nu_{3}!}s_{\eta}^{\left(\nu_{1},\nu_{2},\nu_{3}\right)}\alpha_{1}^{\nu_{1}}\alpha_{2}^{\nu_{2}}\alpha_{3}^{\nu_{3}}+o\left(\left|\alpha\right|^{n}\right),
\]
where $\left|\alpha\right|=\left|\alpha_{1}\right|+\cdots+\left|\alpha_{k}\right|$.
Some properties of cumulants are summarised in \citet[page 19]{Brillinger75}.

The bispectrum is defined as
\[
f_{ijk}^{(3)}\left(\lambda_{1},\lambda_{2}\right)=\left(\frac{1}{2\pi}\right)^{2}\sum_{\alpha_{2}=-\infty}^{\infty}\sum_{\alpha_{1}=-\infty}^{\infty}c_{ijk}\left(\alpha_{1},\alpha_{2}\right)e^{-i\left(\lambda_{1}\alpha_{1}+\lambda_{2}\alpha_{2}\right)},\ \lambda_{1},\lambda_{2}\in\left[-\pi,\pi\right]^{2},\ i,j,k\in\left\{ 1,\ldots,n\right\} ,
\]
where the autocumulant function of order $3$ of $\left(y_{t}\right)$
is defined as $c_{ijk}\left(t_{1},t_{2}\right)=s_{\left(y_{i,t_{1}},y_{j,t_{2}},y_{k,0}\right)}^{\left(1,1,1\right)}$.
It follows from \citet[Theorem 2.8.1, page 34]{Brillinger75} that
the bispectral density of a transformation $k\left(z\right)$, satisfying
a summability condition, of the $n$-dimensional i.i.d. process $\left(\varepsilon_{t}\right)$
is of the form
\[
f_{ijk}^{(3)}\left(\lambda_{1},\lambda_{2}\right)=\left(\frac{1}{2\pi}\right)^{2}\sum_{\gamma=1}^{n}k_{k\gamma}\left(e^{i\left(\lambda_{1}+\lambda_{2}\right)}\right)\sum_{\beta=1}^{n}k_{j\beta}\left(e^{-i\lambda_{2}}\right)\sum_{\alpha=1}^{n}k_{i\alpha}\left(e^{-i\lambda_{1}}\right)S_{\alpha\beta\gamma}
\]
where $S_{ijk}=s_{\left(\varepsilon_{i,t},\varepsilon_{j,t},\varepsilon_{k,t}\right)}^{\left(1,1,1\right)}$.
The objective function in \citet{velasco2020identification} is based
on a vectorised version of $f^{(3)}\left(\lambda_{1},\lambda_{2}\right)$
(as well as the spectral density and the higher order spectral density
of order 4) and compares the true transfer function to its Blaschke
transformed versions. The vectorisation is dervied in \citet{Jammalamadaka_rao_terdik06}.

Unlike this article, \citet{velasco2020identification} does not call
upon the results by \citet{chanho04} for identification of linear
processes satisfying weak summability conditions. The additional structure
of his SVARMA model allows to weaken some assumptions of \citet{chanho04}
such that they are still sufficient for identification. However, \citet{velasco2020identification}
requires the existence of moments while this is not necessary in Assumption
\ref{assu:components_identical}.\footnote{For the essential structure of the proof see \citet{cheng1992} who
treats the univariate case.}

Theorem 4 of \citet{chanho04} is based on Assumption \ref{assu:components_cumu}
and uses higher order cumulants and spectra (and also results by \citet{Jammalamadaka_rao_terdik06})
similarly to \citet{velasco2020identification}. Therefore, it seems
appropriate to compare the conclusions and premises of Theorem 2 in
\citet{velasco2020identification} and Theorem 4 in \citet{chanho04}
more closely. Inspecting the proof of Theorem 4 in \citet{chanho04},
it is easy to see that only stationarity and independence up to order
$4$ would be required as is the case in Assumption 3(4) in \citet{velasco2020identification}.
Following the steps of the proof of Theorem 4 in \citet{chanho04}
and taking the VARMA model structure into account thus leads to Theorem
2 in \citet{velasco2020identification}. However, Theorem 1 in \citet{velasco2020identification}
seems to be slightly stronger than Theorem 7 in \citet{chanho04}
which treats the case where the components of the inputs $\varepsilon_{t}$
are necessarily dependent. While Assumption 2(3) in \citet{velasco2020identification}
implies Assumption D1 in \citet{chanho04}, there seems to be no equivalent
of Assumption D2 in \citet{chanho04} necessary in \citet{velasco2020identification}.

\subsection{Static Identifiability Problem: Choosing a Unique Permutation and
Scaling}

In this section, we describe how to pick one particular permutation
and scaling from the class of observational equivalence described
in the previous section. In order to do this, we describe different
identification schemes, i.e. rules for choosing a particular permutation
and scaling of the matrix $B$ such that $\left(B,\Sigma\right)$
is unique.

We start by repeating an identification scheme presented in \citet{LMS_svarIdent16}
(which are in turn based on \citet{IlmonenPaindaveine11} and \citet{HallinMehta15}).
This identification scheme is convenient for deriving asymptotic properties
and consists firstly of scaling all columns of $B$ such that their
norm is equal to one; secondly, of permutating the columns such that
the absolute value of each diagonal element is larger than the absolute
value of all elements in the same row with a higher column index;
and finally, of scaling all columns of $B$ such that the diagonal
elements are equal to one. This results in an identifiable pair $\left(B,\Sigma\right)$
where $\Sigma$ is a diagonal matrix containing the (positive) standard
deviations of the component densities.

The above transformation exists not on the whole parameter space but
only on a topologically large set in the parameter space. For details,
see Proposition 2 in \citet{LMS_svarIdent16} which includes an example
of a matrix for which the above identification schemes are not defined.
A different identification scheme, similar to the one in \citet{ChenBickel05}
on page 3626, does not exclude any non-singular matrix $B$ and is
defined by the following transformations. Firstly, the columns of
$B$ are scaled to have norm equal to one. Secondly, in each column,
the element with the largest absolute value is made positive. Finally,
the columns are ordered according to $\prec$ such that $c\prec d$
for two columns $c,d$ of $B$ if and only if there exists a $k\in\left\{ 1,\ldots,n\right\} $
such that $c_{k}<d_{k}$ and $c_{j}=d_{j}$ for all $j\in\left\{ 1,\ldots,k-1\right\} $.
This results in a matrix $B$ which incorporates the scalings such
that $\Sigma$ in the pair $\left(B,\Sigma\right)$ is equal to the
identity matrix.

The important takeaway of this section is that the observationally
equivalent points in the parameter space are discrete and therefore
the information matrix is non-singular at each of these discrete points.
In practice, a particular signed permutation needs to be chosen by
the researcher in order to label the shocks. Now that we have firstly
obtained a discrete set of observationally equivalent SVARMA systems
and secondly provided different rules to select a unique representative,
we may proceed to local ML estimation of the true underlying parameter.

\subsection{Further Remarks}
\begin{rem}
When the parameter space for given $\left(p,q\right)$ is not separated
into different fundamentalness regimes, the dynamic and static identifiability
problem cannot be treated independently. Indeed, for transfer functions
$k(z)=a(z)^{-1}b(z)B$ satisfying the assumptions of Section \ref{sec:Model},
it holds that $k_{0}k_{0}'$ is maximal when all zeros of $b(z)$
are outside the unit circle \citep[Theorem 4.2, page 60]{Rozanov67}.
This is a consequence of the fact that the Blaschke factor $b_{\alpha}(z)=\frac{1-\bar{\alpha}z}{z-\alpha}$
which mirrors a zero at $\alpha$ with $\left|\alpha\right|>1$ inside
the unit circle has absolute value smaller than one when evaluated
at $z=0$. Intuitively, this is due to the fact that whenever there
are zeros of the MA polynomial matrix inside the unit circle, the
information space of the agents is strictly larger than the information
space of the outside observer \citep[page 86]{HansenSargent91_2difficulties}.
\end{rem}

\begin{rem}
The connection between the identifiability problem and consistent
estimability - and in particular the importance of identifiability
analysis for the construction of estimators - has been analysed in
\citet{DeistlerSeifert78}. One could argue that the solution to the
identifiability is the more important part of the analysis and that
the construction of an MLE is special in the sense that it is merely
one of many possible estimation procedures. In particular, the main
difference between the analysis of the MLE for the univariate case
of possibly non-invertible ARMA models in \citet{LiiRosenblatt92,LiiRosenblatt96}
and the MLE for the multivariate case treated in the next section
is covered by the identifiability analysis.
\end{rem}
\begin{rem}
\citet{velasco2020identification} allows for zeros inside the unit
circle (except for zeros at zero) of the AR polynomial in the case
of a trivial MA part, i.e. $q=0$. When $q>0$, mirroring poles of
the canonical spectral factor in VARMA representation inside the unit
circle does not seem possible with the approach based on Blaschke
matrices in the parametrisation of rational transfer functions used
by \citet{velasco2020identification}. While the WHF approach in this
article focuses on causal SVARMA models only, it can be extended to
the case where one only requires the rational spectral density has
no zeros and no poles on the unit circle. Note that every rational
transfer function without zeros or poles on the unit circle admits
a WHF.
\end{rem}

\section{\label{sec:MLE}Maximum Likelihood Estimation}

In this section, we treat ML estimation of \eqref{eq:system} in the
parametrisation derived in Theorem \ref{thm:whf} and Corollary \ref{cor:unique_whf}.
Whereas the essential part of this article is the identifiability
analysis of the WHF and the implied non-singularity of the information
matrix of the MLE, the asymptotic theory is standard. We prove consistency
and asymptotic normality on a compact subset of the parameter space
for given integer-valued parameters $\left(p,q,\kappa,k\right)$.
The proof follows the basic structure of the classical consistency
and asymptotic normality proofs \citep[Chapter 3, Chapter 8]{poetpruch97}.
The main ingredient is a uniform law of large numbers (ULLN) for the
(second partial derivatives of the) log-likelihood function which
converges uniformly towards a non-stochastic asymptotic counterpart.
Similarly to the univariate asymptotic analysis in \citet{LiiRosenblatt92,LiiRosenblatt96}
and \citet[Chapter 8]{Rosenblatt00}, we use a Lipschitz-type condition
on the component densities for verifying the ULLNs \citep{andrews87ulln}.
The derivations of the analytic formulae for the score, and the information
matrix are straight-forward but tedious and therefore delegated to
the Appendix. Except for the fact that the polynomial matrices $p(z)$
and $g(z)=s(z)f(z)$ do not commute and the properties of the WHF,
the formulae are similar to the ones in the univariate case in \citep{LiiRosenblatt92,LiiRosenblatt96}.

In the remainder of this section, we firstly describe the (approximate)
log-likelihood function, its asymptotic counterpart, and the parameter
space. Secondly, we state high-level conditions for consistency and
asymptotic normality in order to separate the essential ideas from
technicalities. Lastly, we provide low-level continuity, differentiability,
and integrability assumptions on the component densities which are
sufficient for the high-level conditions.\textcolor{red}{}

\subsection{Parameter Space and (Approximate) Log-Likelihood Function}

We start by describing the parameter space over which we optimise
the log-likelihood function. Subsequently, we make assumptions on
the densities of the components of $\varepsilon_{t}$ in order to
provide explicit expressions for the individual contributions to the
standardised log-likelihood function and its partial derivatives.

For given integer valued parameters $\left(p,q,\left(\kappa,k\right)\right)$,
we vectorise the system parameters, i.e. the ones in $\left(a(z),p(z),f(z)\right)$,
in column-major order. The AR parameters are vectorised as $\tau_{1}=vec\left(a_{1},\ldots,a_{p}\right),$
the ``stable'' MA parameters for $\left(\kappa,0\right)$ as $\tau_{2}=vec\left(p_{1},\ldots,p_{q-\kappa}\right)$
and for $\left(\kappa,k\right),\ k>0,$ as
\[
\tau_{2}=vec\left(\begin{pmatrix}I_{k} & 0_{k\times(n-k)}\\
p_{0,21} & I_{n-k}
\end{pmatrix},\begin{pmatrix}p_{1,11} & 0_{k\times(n-k)}\\
p_{1,21} & p_{1,22}
\end{pmatrix},\ldots,p_{\kappa-1},\begin{pmatrix}0_{n\times k} & p_{\kappa,\left[\bullet,k+1:n\right]}\end{pmatrix}\right).
\]
It turns out that it is more convenient to parametrise the ``unstable''
MA parameters in 
\begin{align*}
g(z) & =s(z)f(z)\\
 & =\begin{pmatrix}f_{\kappa+1,[1:k,\bullet]}\\
f_{\kappa,[k+1:n,\bullet]}
\end{pmatrix}+\begin{pmatrix}f_{\kappa,[1:k,\bullet]}\\
f_{\kappa-1,[k+1:n,\bullet]}
\end{pmatrix}z+\cdots+\begin{pmatrix}f_{1,[1:k,\bullet]}\\
f_{0,[k+1:n,\bullet]}
\end{pmatrix}z^{\kappa}+\begin{pmatrix}f_{0,[1:k,\bullet]}\\
0_{(n-k)\times n}
\end{pmatrix}z^{\kappa+1}
\end{align*}
rather than the ones in $f(z)$ directly. Of course, they are in a
one-to-one relation and can be easily obtained from each other, whenever
necessary. In the natural parametrisation, $f_{0}$ is restricted
to be the identity matrix. If one imposes the additional assumption
that $b_{0}$ is non-singular and normalises it to the identity matrix,
this entails for $g_{0}$ that it is equal to $\left(\begin{smallmatrix}I_{k} & 0_{k\times(n-k)}\\
-p_{0,21} & I_{n-k}
\end{smallmatrix}\right)$ for $k>0$, and the identity matrix otherwise. The parameters in
$g(z)$ are vectorised as 
\[
\tau_{3}=vec\left(g_{0},g_{1},\ldots,g_{\kappa},\begin{pmatrix}g_{\kappa+1,[1:k,\bullet]}\\
0_{(n-k)\times n}
\end{pmatrix}\right)
\]
in the case $\left(\kappa,k\right)$, $k>0$ and as $\tau_{3}=vec\left(g_{1},\ldots,g_{\kappa}\right)$
if $k=0$. Again, $\left(\begin{smallmatrix}g_{\kappa+1,[1:k,\bullet]}\\
g_{\kappa,[k+1:n,\bullet]}
\end{smallmatrix}\right)=I_{n}$ in the natural parametrisation and $g_{0}=\left(\begin{smallmatrix}I_{k} & 0_{k\times(n-k)}\\
-p_{0,21} & I_{n-k}
\end{smallmatrix}\right)$ when $b_{0}=I_{n}$ is required.

Obviously, not all parameters in $\tau'=\left(\tau_{1}',\tau_{2}',\tau_{3}'\right)$
are free. As can be easily seen from the vectorisations above, there
are $n(n-1)+kn$ zero-restrictions and $n$ one-restrictions in $\tau_{2}$.
Regarding restrictions on $\tau_{3}$, there are $n\left(n-1\right)+kn$
zero restrictions and $n$ one restrictions in the natural parametrisation,
whereas in the parametrisation where one additional requires $b_{0}=I_{n}$,
there are $n^{2}$ zero and one restrictions on $\tau_{3}$ and $k(n-k)$
restrictions between the parameters in $\tau_{2}$ and $\tau_{3}$.
We represent these restrictions as $R\tau=r$ where $R$ is of full
row rank and of dimension $3n^{2}\times n_{\tau}$, where $n_{\tau}=n^{2}\left(p+q+3\right)$.
Note that the number of free system parameters is therefore $n^{2}\left(p+q\right)$
and does not depend on $\left(\kappa,k\right)$ (and thus on the fundamentalness
regime).

The (free) parameters pertaining to the underlying economic shocks
are vectorised and summarised in
\begin{assumption}
\label{assu:paramSpace}The true parameter value $\theta_{0}$ belongs
to the permissible parameter space $\Theta=\Theta_{\tau}\times\Theta_{\beta}\times\Theta_{\sigma}\times\Theta_{\lambda}=\Theta_{\tau}\times\Theta_{\gamma},$
where
\begin{enumerate}
\item $\Theta_{\tau}$ with $\Theta_{\tau}\subseteq\mathbb{R}^{n^{2}\left(p+q\right)}$
is such that conditions \eqref{eq:stability}, \eqref{eq:no_unitcircle_zeros},
the coprimeness assumption and the full rank assumption on $\left(a_{p},b_{q}\right)$
are satisfied, and
\item $\Theta_{\beta}=vecd\text{°}\left(\mathcal{B}\right)=\left\{ \beta\in\mathbb{R}^{n(n-1)}\,|\,\beta=vecd\text{°}\left(B\right)\text{ for some }B\in\mathcal{B}\right\} $.
The vector $\beta$ collects the off-diagonal elements of $B$.
\item For the scalings, $\Theta_{\sigma}=\mathbb{R}_{+}^{n}$ holds, and
\item for the additional parameters appearing in the component densities,
we have $\Theta_{\lambda}=\Theta_{\lambda_{1}}\times\cdots\times\Theta_{\lambda_{n}}\subseteq\mathbb{R}^{d}$
with $\Theta_{\lambda_{i}}\subseteq\mathbb{R}^{d_{i}}$ open for every
$i\in\left\{ 1,\ldots,n\right\} $ and $d=d_{1}+\cdots+d_{n}$.
\end{enumerate}
\end{assumption}
We also introduce the non-singleton compact and convex subset $\Theta_{0}=\Theta_{0,\tau}\times\Theta_{0,\gamma}$
of the interior of $\Theta$ which contains the true parameter value
$\theta_{0}$.

Regarding the component densities of the i.i.d. shock process $\left(\varepsilon_{t}\right)$,
we state
\begin{assumption}
\label{assu:densities}For each $i\in\left\{ 1,\ldots,n\right\} $
the distribution of the error term $\varepsilon_{i,t}$ has a (Lebesgue)
density $f_{i,\sigma_{i}}\left(x;\lambda_{i}\right)=\sigma_{i}^{-1}f_{i}\left(\sigma_{i}^{-1}x;\lambda_{i}\right)$
which may also depend on a parameter vector $\lambda_{i}\in\mathbb{R}^{d_{i}}$.
\end{assumption}
The family of skewed generalised t-distributions (SGT) is parametrised
by three parameters $\left(\mathfrak{l},\mathfrak{p},\mathfrak{q}\right)$
when the mean and variance are required to be zero and one) and is
sufficiently rich for our requirements \citep{theodossiou98sgt,davis16sgt_pkg}.
The parameter $\mathfrak{l}\in\left(-1,1\right)$ parametrises the
skewness of the distribution and $\left(\mathfrak{p},\mathfrak{q}\right)$
parametrise the kurtosis. The (skewed) Laplace, Cauchy, and t-distribution
are obtained for $\left(\mathfrak{p},\mathfrak{q}\right)$ equal to
$\left(1,\infty\right)$, $\left(2,\frac{1}{2}\right)$, and $\left(2,r\right)$
where $r$ denotes the degrees of freedom of the t-distribution, respectively.
The normal distribution, for $\left(\mathfrak{l},\mathfrak{p},\mathfrak{q}\right)=\left(0,2,\infty\right)$,
and the uniform distribution, for $\mathfrak{p}\rightarrow\infty$,
are contained in this family as limit cases. Moreover, the existence
of moments up to a certain order can be ensured by conditions on the
parameters of the distribution. In particular the expectation and
the variance exist for $\mathfrak{pq}>1$ and $\mathfrak{pq}>2$,
respectively.

The (standardised) approximate log-likelihood function is defined
as
\begin{align}
L_{T}\left(\theta,y_{1},\ldots,y_{T}\right) & =\frac{1}{T}\sum_{t=1}^{T}l_{t}\left(\varepsilon_{t}(\theta),\theta\right)\label{eq:likelihood}
\end{align}
where 
\begin{align}
l_{t}\left(\varepsilon_{t}(\theta),\theta\right) & =\sum_{i=1}^{n}\log\left[f_{i}\left(\sigma_{i}^{-1}\iota_{i}^{'}B\left(\beta\right)^{-1}u_{t}\left(\theta\right);\lambda_{i}\right)\right]\label{eq:likelihood_individual}\\
 & \quad-\log\left\{ \left|\det\left(f_{0}\right)\right|\right\} -\log\left\{ \left|\det\left(B\left(\beta\right)\right)\right|\right\} -\sum_{i=1}^{n}\log\left(\sigma_{i}\right),\nonumber 
\end{align}
in which $\iota_{i}$ is the unit column-vector with a one at the
$i$-th position, $u_{t}\left(\theta\right)=B\varepsilon_{t}\left(\theta\right)$,
and 
\begin{align*}
\varepsilon_{t}(\theta) & =B^{-1}f(z)^{-1}s(z)^{-1}p(z)^{-1}a(z)y_{t}\\
 & =\sum_{\substack{j=-\infty\\
1\leq t\leq T
}
}^{\infty}w_{j}y_{t-j}=\sum_{s=1}^{T}w_{t-s}y_{s}.
\end{align*}

As additional device for the proof that $L_{T}\left(\theta,y_{1},\ldots,y_{T}\right)$
converges uniformly on $\Theta_{0}$, we define 
\[
L_{T}^{\infty}\left(\theta,\left(y_{t}\right)_{t\in\mathbb{Z}}\right)=\frac{1}{T}\sum_{t=1}^{T}l_{t}\left(\tilde{\varepsilon}_{t}(\theta),\theta\right)
\]
where $\varepsilon_{t}(\theta)$ is replaced with $\tilde{\varepsilon}_{t}(\theta)=\sum_{j=-\infty}^{\infty}w_{j}y_{t-j}.$
The asymptotic counterpart, to which $L_{T}\left(\theta,y_{1},\ldots,y_{T}\right)$
converges under appropriate conditions outlined below and which is
minimised at the true $\theta_{0}$, is $L\left(\theta\right)=\mathbb{E}\left(l_{t}\left(\tilde{\varepsilon}_{t}(\theta),\theta\right)\right)$.

\subsection{High-Level Arguments for Consistency and Asymptotic Normality}

The following propositions are based on \citet[Chapter 3, Chapter 8]{poetpruch97}
and \citet{poet11} and summarise the basic ingredients necessary
to prove consistency and asymptotic normality of the MLE.
\begin{prop}
\label{prop:generic_consistency}Let $\Theta\subseteq\mathbb{R}^{q}$
be compact and let $L:\Theta\rightarrow\mathbb{R}$ be a non-random
continuous function which is uniquely minimised at $\theta_{0}$.
If $L_{T}\left(\theta,y_{1},\ldots,y_{T}\right)$ satisfies a ULLN,
i.e. $\sup_{\Theta}\left|L_{T}\left(\theta,y_{1},\ldots,y_{T}\right)-L(\theta)\right|\xrightarrow{T\rightarrow\infty}0$,
then the sequence of minimisers $\left(\hat{\theta}_{T}\right)$ of
$L_{T}\left(\theta,y_{1},\ldots,y_{T}\right)$ converges in probability
to $\theta_{0}$.
\end{prop}
Similarly to \citet[page 15]{LiiRosenblatt96}, it can be shown that
$\sup_{\Theta_{0}}\left|L_{T}\left(\theta,y_{1},\ldots,y_{T}\right)-L_{T}^{\infty}\left(\theta,\left(y_{t}\right)_{t\in\mathbb{Z}}\right)\right|\xrightarrow{T\rightarrow\infty}0$
under a Lipschitz condition, see Assumption \ref{assu:MLE}.5 below,
on the derivative of the component densities and due to the rational
structure of our model \citep[page 17]{deistler75}. The ULLN $\sup_{\Theta_{0}}\left|L_{T}^{\infty}\left(\theta,\left(y_{t}\right)_{t\in\mathbb{Z}}\right)-L(\theta)\right|\xrightarrow{T\rightarrow\infty}0$,
where $l_{t}\left(\tilde{\varepsilon}_{t}(\theta),\theta\right)$
depends on $t$ only through $\tilde{\varepsilon}_{t}(\theta)$, follows
from the generic ULLN in \citet{andrews87ulln} from verifying the
first moment continuity condition \citep[Assumption (A3)]{andrews87ulln}
as well as a local LLN \citep[Assumption (A2)]{andrews87ulln} for
bracketing functions $\sup_{\Theta_{0}}l_{t}\left(\tilde{\varepsilon}_{t}(\theta),\theta\right)$
and $\inf_{\Theta_{0}}l_{t}\left(\tilde{\varepsilon}_{t}(\theta),\theta\right)$.
These assumptions in \citet{andrews87ulln} in turn follow from the
ergodic theorem and under appropriate boundedness assumptions (which
are implied by Assumption \citet[Assumption (a.5)]{GourierouxMR_svarma19}).\textcolor{red}{}

Regarding asymptotic normality, we have
\begin{prop}
\label{prop:generic_asynormal}Let $L_{T}\left(\theta,y_{1},\ldots,y_{T}\right)$
be twice continuously differentiable on an open subset $U$, containing
$\theta_{0}$, of the compact set $\Theta\subseteq\mathbb{R}^{q}$.
Let $\sqrt{T}\left.\frac{\partial L_{T}\left(\theta,y_{1},\ldots,y_{T}\right)}{\partial\theta}\right|_{\theta=\theta_{0}}$
satisfy a CLT such that it converges in distribution to $\mathcal{N}\left(0,B\right)$.
Furthermore, assume that $\sqrt{T}\left.\frac{\partial^{2}L_{T}\left(\theta,y_{1},\ldots,y_{T}\right)}{\partial\theta\partial\theta'}\right|_{\theta=\theta_{0}}$
converges uniformly in probability on a compact subset $K$ of $U$
with positive radius and center $\theta_{0}$ to a non-random matrix
$A(\theta)$ which is non-singular evaluated at $\theta_{0}$ and
continuous on $K$. It follows that $\sqrt{T}\left(\hat{\theta}_{T}-\theta_{0}\right)\xrightarrow{d}\mathcal{N}\left(0,A\left(\theta_{0}\right)^{-1}BA\left(\theta_{0}\right)^{-1}\right)$
holds, where $\left(\hat{\theta}_{T}\right)$ satisfies $\sqrt{T}\left.\frac{\partial L_{T}\left(\theta,y_{1},\ldots,y_{T}\right)}{\partial\theta}\right|_{\theta=\hat{\theta}_{T}}=0$
and converges in probability to $\theta_{0}$.
\end{prop}
Except for the multivariate nature, this follows essentially from
\citet{LiiRosenblatt92,LiiRosenblatt96}.\textcolor{red}{}

Again, the main ingredient is the verification ULLN for the second
partial derivative using a Lipschitz-like condition. The CLT (in which
$B=\mathbb{E}\left(\frac{\partial L_{T}\left(\theta_{0},y_{1},\ldots,y_{T}\right)}{\partial\theta}\frac{\partial L_{T}\left(\theta_{0},y_{1},\ldots,y_{T}\right)}{\partial\theta'}\right)$
and $A\left(\theta_{0}\right)=\mathbb{E}\left(\frac{\partial^{2}L_{T}\left(\theta_{0},y_{1},\ldots,y_{T}\right)}{\partial\theta\partial\theta'}\right)$)
follows from the CLT for $m$-dependent processes \citep[Theorem 6.4.2, page 213]{BrockwellDavis87}
and Bernstein's Lemma \citep[page 242]{Hannan70}. Note that for the
CLT, the requirement that a sequence of minimisers $\left(\hat{\theta}_{T}\right)$
of $L_{T}\left(\theta,y_{1},\ldots,y_{T}\right)$ converges in probability
to $\theta_{0}$ can be relaxed to requiring that $\frac{\partial L_{T}\left(\hat{\theta}_{T},y_{1},\ldots,y_{T}\right)}{\partial\theta}$
converges (in probability) faster to zero than $\frac{1}{\sqrt{T}}$.
This is the approach taken in \citet{LiiRosenblatt92,LiiRosenblatt96}
who show the existence of such a consistent root and focus directly
on asymptotic normality, presumably due to the close connection of
consistency and identifiability.

\subsection{Low-Level Assumptions and Scores}

The following assumptions are similar to \citet{LiiRosenblatt92,LMS_svarIdent16}
and satisfied in particular by the SGT family of densities for appropriate
parameters $\left(\mathfrak{l},\mathfrak{p},\mathfrak{q}\right)$
and linear combinations thereof (including Gaussian mixtures).
\begin{assumption}
\label{assu:MLE}The following conditions hold for $i\in\left\{ 1,\ldots,n\right\} $.
\begin{enumerate}
\item For all $x\in\mathbb{R}$ and all $\lambda_{i}\in\Theta_{0,\lambda_{i}},\ f_{i}\left(x;\lambda_{i}\right)>0$
and $f_{i}\left(x;\lambda_{i}\right)$ is twice continuously differentiable
with respect to $\left(x;\lambda_{i}\right)$.
\item The function $f_{i,x}\left(x;\lambda_{i,0}\right)$ is integrable
with respect to x, and $\int f_{i,x}\left(x;\lambda_{i,0}\right)dx=0$.
Moreover, $\int\sup_{\lambda_{i}\in\Theta_{0,\lambda_{i}}}\left\Vert f_{i,\lambda_{i}}\left(x;\lambda_{i}\right)\right\Vert dx=\mathbb{E}\left(\frac{\sup_{\lambda_{i}\in\Theta_{0,\lambda_{i}}}\left\Vert f_{i,\lambda_{i}}\left(x;\lambda_{i}\right)\right\Vert }{f_{i}\left(x;\lambda_{i,0}\right)}\right)<\infty$.
\item The functions $f_{i,xx}\left(x;\lambda_{i,0}\right)$ and $f_{i,x\lambda_{i}}\left(x;\lambda_{i,0}\right)$
are integrable with respect to $x$, i.e., 
\[
\int\left|f_{i,xx}\left(x;\lambda_{i,0}\right)\right|dx<\infty\ and\ \int\left\Vert f_{i,x\lambda_{i}}\left(x;\lambda_{i,0}\right)\right\Vert dx<\infty,
\]
and furthermore $\int\sup_{\lambda_{i}\in\Theta_{0,\lambda_{i}}}\left\Vert f_{i,\lambda_{i}\lambda_{i}}\left(x;\lambda_{i}\right)\right\Vert dx<\infty$
holds.
\item For all $x\in\mathbb{R}$
\[
x^{2}\frac{f_{i,x}^{2}\left(x;\lambda_{i}\right)}{f_{i}^{2}\left(x;\lambda_{i}\right)}\ and\ \frac{\left\Vert f_{i,\lambda_{i}}\left(x;\lambda_{i}\right)\right\Vert ^{2}}{f_{i}^{2}\left(x;\lambda_{i}\right)}
\]
are dominated by $c_{1}\left(1+\left|x\right|^{c_{2}}\right)$ with
$c_{1},c_{2}\geq0$ and $\int\left|x\right|^{c_{2}}f_{i}\left(x;\lambda_{i,0}\right)dx=\mathbb{E}\left(\left|x\right|^{c_{2}}\right)<\infty$.
\item For $u(x)=\frac{f_{i,x}(x;\lambda_{i})}{f_{i}(x;\lambda_{i})}$ or
$u(x)=\frac{\partial}{\partial x}\left(\frac{f_{i,x}(x;\lambda_{i})}{f_{i}(x;\lambda_{i})}\right)$,
the generalised Lipschitz condition $\left|u\left(x+h,\lambda_{i}\right)-u\left(x+h,\lambda_{i}\right)\right|\leq A\left[\left(1+\left|x\right|^{k}\right)\left|h\right|+\left|h\right|^{l}\right]$,
and $\mathbb{E}\left(\left|x\right|^{2+k+l}\right)<\infty$ hold.
\end{enumerate}
\end{assumption}

The expressions for the partial derivatives of the individual contributions
to the log-likelihood function are given as 
\begin{align*}
\frac{\partial l_{t}\left(\theta\right)}{\partial\tau_{1}} & =-x_{b,t-1}\left(\theta\right)B'\left(\beta\right)^{-1}\Sigma^{-1}e_{x,t}\left(\theta\right)\\
\frac{\partial l_{t}\left(\theta\right)}{\partial\tau_{2}} & =-\left(\left[f(z)^{-1}s(z)^{-1}p(z)^{-1}\right]\left[w_{g,t-1}'\left(\theta\right)\otimes I_{n}\right]\right)'B'\left(\beta\right)^{-1}\Sigma^{-1}e_{x,t}\left(\theta\right)\\
\frac{\partial l_{t}\left(\theta\right)}{\partial\tau_{3}} & =-\left(\left[f(z)^{-1}s(z)^{-1}p(z)^{-1}\right]\left[w_{p,t-1}'\left(\theta\right)\otimes I_{n}\right]\right)'B'\left(\beta\right)^{-1}\Sigma^{-1}e_{x,t}\left(\theta\right)-\frac{\partial vec\left(f_{0}\right)}{\partial\tau_{3}}vec\left(f_{0}'^{-1}\right)\\
\frac{\partial l_{t}\left(\theta\right)}{\partial\beta} & =-H'\sum_{i=1}^{q}\left(B\left(\beta\right)^{-1}u_{t-i}\left(\theta\right)\otimes b_{i}'B'\left(\beta\right)^{-1}\Sigma^{-1}e_{x,t}\left(\theta\right)\right)\\
 & \qquad-H'\left(B\left(\beta\right)^{-1}u_{t}\left(\theta\right)\otimes B'\left(\beta\right)^{-1}\Sigma^{-1}e_{x,t}\left(\theta\right)\right)\\
 & \qquad-H'vec\left(B'\left(\beta\right)^{-1}\right)\\
\frac{\partial}{\partial\sigma}l_{t}\left(\theta\right) & =-\Sigma^{-2}\left[e_{x,t}\left(\theta\right)\odot\varepsilon_{t}\left(\theta\right)+\sigma\right]\\
\frac{\partial}{\partial\lambda}l_{t}\left(\theta\right) & =e_{\lambda,t}\left(\theta\right)
\end{align*}

where $x_{b,t-1}'=\left[f(z)^{-1}z^{-\kappa}p(z)^{-1}\right]\left[x_{t-1}'\otimes I_{n}\right]$,
$x'_{t-1}=\left(y'_{t-1},\ldots,y'_{t-p}\right)$, 
\begin{align*}
w_{g,t-1}' & =\left(g(z)u_{1,t-1}(\theta),\ldots,g(z)u_{n,t-1}(\theta)|\cdots|g(z)u_{1,t-(q-\kappa)}(\theta),\ldots,g(z)u_{n,t-(q-\kappa)}(\theta)\right),\\
w_{p,t-1}' & =\left(p(z)u_{1,t-1}(\theta),\ldots,p(z)u_{n,t-1}(\theta)|\cdots|p(z)u_{1,t-\kappa}(\theta),\ldots,p(z)u_{n,t-\kappa}(\theta)\right),
\end{align*}
the matrix $H\in\mathbb{R}^{n^{2}\times n(n-1)}$ consisting of zeros
and ones is implicitly defined by $vec\left(B(\beta)\right)=H\beta+vec\left(I_{n}\right)$
for $B$ in $\mathcal{B}$.

The other main differences in the partial derivatives of the log-likelihood
function compared to the invertible Gaussian case are the appearance
of $f(z)$ and $g(z),$ the term $\log\left\{ \left|\det\left(f_{0}\right)\right|\right\} $,
and the fact that the expressions
\[
e_{i,x,t}(\theta)=\frac{\partial}{\partial x}\log\left[f_{i}\left(\sigma_{i}^{-1}\iota_{i}^{'}B\left(\beta\right)^{-1}u_{t}\left(\theta\right);\lambda_{i}\right)\right]=\frac{f_{i,x}\left(\sigma_{i}^{-1}\varepsilon_{i,t}\left(\theta\right);\lambda_{i}\right)}{f_{i}\left(\sigma_{i}^{-1}\varepsilon_{i,t}\left(\theta\right);\lambda_{i}\right)}
\]
and 
\[
e_{i,\lambda_{i},t}(\theta)=\frac{\partial}{\partial\lambda_{i}}\log\left[f_{i}\left(\sigma_{i}^{-1}\iota_{i}^{'}B\left(\beta\right)^{-1}u_{t}\left(\theta\right);\lambda_{i}\right)\right]=\frac{f_{i,\lambda}\left(\sigma_{i}^{-1}\varepsilon_{i,t}\left(\theta\right);\lambda_{i}\right)}{f_{i}\left(\sigma_{i}^{-1}\varepsilon_{i,t}\left(\theta\right);\lambda_{i}\right)},
\]
with $f_{i,x}\left(x;\lambda_{i}\right)=\frac{\partial}{\partial x}f_{i}\left(x;\lambda_{i}\right)$
and $f_{i,\lambda_{i}}\left(x;\lambda_{i}\right)=\frac{\partial}{\partial\lambda_{i}}f_{i}\left(x;\lambda_{i}\right)$
do not simplify as in the Gaussian case (compare the terms $C_{1},C_{2}$
in \citet[page 277]{LiiRosenblatt92} and $\tilde{I},\tilde{J}$ in
\citet[Chapter 8]{Rosenblatt00}). Evaluated at the truth, i.e. $\theta=\theta_{0}$,
we have that $\varepsilon_{i,t}\left(\theta_{0}\right)=\varepsilon_{i,t}$
and 
\[
e_{i,x,t}=e_{i,x,t}(\theta_{0})=\left.\frac{\partial}{\partial x}\log\left[f_{i}\left(\sigma_{i}^{-1}\iota_{i}^{'}B\left(\beta\right)^{-1}u_{t}\left(\pi\right);\lambda_{i}\right)\right]\right|_{\theta=\theta_{0}}=\frac{f_{i,x}\left(\sigma_{i}^{-1}\varepsilon_{i,t};\lambda_{i,0}\right)}{f_{i}\left(\sigma_{i,0}^{-1}\varepsilon_{i,t};\lambda_{i,0}\right)}.
\]

In combination, these assumptions allow us to prove
\begin{thm}
\label{thm:MLE}Under Assumptions \ref{assu:paramSpace}, \ref{assu:densities},
\ref{thm:MLE}, and one of Assumptions \ref{assu:components_identical}
or \ref{assu:components_cumu}, there exists a sequence of maximisers
$\hat{\theta}_{T}$ of \eqref{eq:likelihood} such that $\sqrt{T}\left(\hat{\theta}_{T}-\theta_{0}\right)$
converges in distribution to $\mathcal{N}\left(0,S\right)$, where
\[
S=\begin{pmatrix}I_{0} & R'\\
R & 0
\end{pmatrix}^{-1}\begin{pmatrix}I_{0} & 0\\
0 & 0
\end{pmatrix}\begin{pmatrix}I_{0} & R'\\
R & 0
\end{pmatrix}^{-1}
\]
and $I_{0}=\mathbb{E}\left[l_{\theta,t}\left(\theta_{0}\right)l_{\theta,t}'\left(\theta_{0}\right)\right]$.
\end{thm}
\begin{proof}
The Theorem follows from Propositions \ref{prop:generic_consistency}
and \ref{prop:generic_asynormal} by verification of the high-level
assumptions as outlined above. \textcolor{red}{}
\end{proof}

\subsection{Further Remarks}

\citet{GourierouxMR_svarma19} use similar high level assumptions
albeit without verifying them. In particular, they assume in (a.1)
that the parameter space, to which their Lemma 1 and Proposition 2
is applied, is compact. However, it is well known that the parameter
space for multivariate SVARMA models of the kind in their article
is not compact. Moreover, no attempt is made to verify that their
assumption (a.2), that the Schur decomposition be selected such that
it is continuous for any given non-fundamentalness regime (i.e. number
of MA zeros inside the unit circle), holds. The WHF approach as detailed
in Theorem \ref{thm:whf} and Corollary \ref{cor:unique_whf} provides
a solution for this and could be used to verify (a.2) in \citet{GourierouxMR_svarma19}.
Therefore, \citet{GourierouxMR_svarma19} just formulate a well-known
problem without providing a solution and their Proposition 2, presumably
their main result, is empty.

Proving consistency for the whole parameter space (in contrast to
a compact subset) is significantly more difficult, see \citet[Chapter 4]{HannanDeistler12}.
In this general case, the proof is not conducted in coordinates (which
are only introduced when analysing asymptotic normality in a specific
parametrisation). A succinct description of the difficulties in this
case is given in \citet[Chapter 4.5]{poetpruch97}, see also \citet[Chapter 4.6, page 35, footnote 13]{poetpruch97}.
The non-compactness of the parameter space is dealt with by first
extending the likelihood function to be minimised to a larger space
$\Theta^{**}\supseteq\Theta$ and subsequently showing that one obtains
the same minimiser for optimising the likelihood function on $\Theta$
or on a compact subset $\Theta^{*}\subseteq\Theta^{**}$. Still, the
likelihood function does not converge uniformly to its asymptotic
counterpart without further adjustments. The convergence properties
of the likelihood function for parameter values such that there are
transfer function zeros close to or on the unit circle requires some
more analysis and adjustments \citep{DeistlerPoetscher84,poet87,dahlhauspoet89}.
Note that \citet{GourierouxMR_svarma19} do not analyse the convergence
properties of their exact likelihood function (derived for the univariate
MA(1) case) when there is a zero of the MA polynomial on the unit
circle. \citet{rissanencaines79mle} prove consistency and asymptotic
normality of the MLE (similarly to this article) when the parameter
space is restricted to a compact set and the inputs are i.i.d. Gaussian.

\section{Empirical Application}

In order to illustrate the advantages of the WHF approach compared
to other methods \citep{GourierouxMR_svarma19,velasco2020identification},
we estimate the Blanchard-Quah model \citet{blanchard_quah89}. Furthermore,
we show the potential usefulness of the WHF approach for discriminating
between different fundamentalness regimes for given AR and MA orders
by estimating a standard macroeconomic model (federal funds rate,
unemployment rate, inflation) with and without real exchange rates.
Details are contained in the associated R-package, including another
application to the dataset used in \citet{plagbormoller19SVMA} who
analyses technological news shocks and business cycles in the context
of Bayesian SVMA models.

\subsection{Economic Problem in \citet{blanchard_quah89} and Summary of Existing
Results}

\citet{blanchard_quah89} analyse the impact of demand and supply
shocks on GNP and unemployment. They identify the shocks (the static
shock transmission matrix) by assuming that demand shocks have only
a temporary effect on GNP, while supply shocks have a permanent effect
and estimate a SVAR(8) model for the first differences of the logarithm
of output and the (detrended) unemployment rate in order to obtain
IRFs. In the plots below, the first differences of the logarithm of
output are aggregated. \citet{lippi_reichlin93aer,LippiReichlin94}
criticise that autoregressive models exclude zeros of the transfer
function apriori and argue that \citet{blanchard_quah89}'s VAR(8)
approximates\footnote{While in theory one may approximate SVARMA models (or even ``infinite
VAR models'') by SVAR models, it is well known that the approximation
is bad in many practically relevant cases. This was emphasised in
a macroeconometric context by \citet{Ravenna07} and \citet{PoskittYao17}.
\citet{Ravenna07} decomposes the error when SVARMA models are approximated
by SVAR models into a truncation error and an identification error,
pertaining to the parameters describing the economic shocks. \citet{PoskittYao17}
decompose the truncation error introduced in \citet{Ravenna07} further
into an estimation and approximation error and argue that both are
large for commonly used lag lengths and sample sizes. They conclude
that ``using VAR($n$) may not be justified unless $n$ and {[}the
sample size{]} $T$ are enormous''. Obviously, these errors carry
over to the IRF which is a non-linear transformation of the structural
parameters.} a VARMA(1,1) model with complex-conjugated MA roots \citet[page 325]{LippiReichlin94}.
This implies in particular that for the SVARMA(1,1) model there are
either two MA zeros inside or two MA zeros outside the unit circle.

The approach in \citet{velasco2020identification} results in an invertible
SVARMA(1,1) model with impulse responses similar to the ones by \citet{blanchard_quah89}\footnote{Since the impulse respsonses obtained by \citet{velasco2020identification}
are essentially qualitatively the same as the ones in \citet{blanchard_quah89},
we do not plot them separately.}. The approach in \citet{GourierouxMR_svarma19} results in a SVARMA(4,1)
model with one root inside the unit circle which is not in line with
the results in \citet{LippiReichlin94}. Be that as it may, the impulse
responses obtained in \citet{GourierouxMR_svarma19} are similar to
the ones obtained from the SVARMA-WHF approach in this article, albeit
for a SVARMA(1,2,1,0) model with two MA roots inside the unit circle.
These differences in model selection might be due to the fact that
the data are not informative enough and that it is therefore difficult
to infer the correct maximising parameter vector, see \citet{YaoKamVahid17weakVARMA}.

\subsection{Description of the WHF Approach}

The elegance and simplicity of the SVARMA-WHF approach allows us to
investigate a multitude of different AR and MA orders $p$ and $q$.
For each combination $\left(p,q\right)\in\left\{ 0,\ldots,8\right\} ^{2}$,
we estimate all possible a SVARMA-WHF models $\left(\kappa,k\right)$.
By Theorem \ref{thm:whf} and Corollary \ref{cor:unique_whf}, the
number of free parameters depends only on $\left(p,q\right)$ and
therefore it makes sense to use model selection criteria like AIC
and BIC \citep{burnhamanderson04modelselection,claeskenshjort08model}.
Following the discussion in \citet[Remark 1, page 171]{poet90armaorder},
we use BIC as model selection criterion.

In order to investigate the convergence behaviour of our estimation
procedure, we estimate the model with the Gaussian density\footnote{As soon as a fundamentalness regime is fixed, the estimation problem
is identical to the one for causal invertible VARMA models. In the
latter model, the number of MA roots inside the unit circle is zero
by assumption.}, the Laplace density, and finally with the SGT family of densities.
For each of these densities, we iterate between the BFGS (L-BFGS-B
in the case of the SGT family in order to take the boundedness of
the parameters into account and to ensure that the densities are such
that a certain number of moments exist) and Nelder-Mead algorithm
and use the obtained maximising parameter value as starting value
for the subsequent (family of) densities.

We check the residuals for non-normality using the Jarque-Bera and
Shapiro tests. Moreover, we investigate the independence properties
of the components of the obtained shock processes by analysing their
autocorrelation properties as well as those of their absolute values
and their squares. If the shock processes are independent, the shock
processes as well as their absolute values and their squares should
be uncorrelated. However, formal results regarding the properties
of these test statistics (and afortiori small sample properties) under
non-normality and taking estimation uncertainty into account have
yet to be developed\footnote{We observe that the Ljung-Box test rejects more often when the deviation
from non-normality is remarkable.}. Therefore, we use the output of these test statistics only as a
guiding principle for comparison of similar models.

\subsection{Results}

The model with minimal BIC value and for which the normality tests
reject the $H_{0}$ of normality at level $10\%$ and do not reject
the $H_{0}$ of the Ljung-Box test (with $8$ lags) for each component
of the estimated shocks is the SVARMA(1,2,1,0) model. The next best
ones are the SVARMA(1,3,1,0) and the SVARMA(3,1,1,0) model. Each of
these models feature two MA roots inside the unit circle and imply
similar IRFs.

The optimal parameter values (with their standard deviations in parentheses)
are 
\begin{align*}
\hat{a}(z)=I_{2}+\begin{pmatrix}\underset{(2.61)}{-0.11} & \underset{(2.62)}{-0.28}\\
\underset{(4.92)}{0.49} & \underset{(8.29)}{-1.01}
\end{pmatrix},\quad & \hat{p}(z)=I_{2}+\begin{pmatrix}\underset{(2.62)}{0.07} & \underset{(1.1)}{0.75}\\
\underset{(4.85)}{0.26} & \underset{(1.6)}{0.00}
\end{pmatrix}\\
\hat{f}(z)=I_{2}+\begin{pmatrix}\underset{(2.20)}{0.56} & \underset{(0.610)}{1.67}\\
\underset{(4.61)}{-0.24} & \underset{(1.37)}{-0.30}
\end{pmatrix},\quad & \hat{B}=\begin{pmatrix}\underset{(1.63)}{0.94} & \underset{(1.94)}{0.29}\\
\underset{(4.39)}{-0.22} & \underset{(5.65)}{0.18}
\end{pmatrix}
\end{align*}

and the parameters $\left(\mathfrak{l},\mathfrak{p},\mathfrak{q}\right)$
for the shock densities are $\left(-0.52,1.49,85616\right)$ (standard
deviations $\left(1.57,0.42,0\right)$) and $\left(0.15,1.92,7.89\right)$
(standard deviations $\left(0.67,0.28,0.01\right)$), respectively.

Next, we compare the IRFs of our model with the ones of \citet{blanchard_quah89}
and in \citet{GourierouxMR_svarma19} in Figure 1. Unlike the impulse
responses in \citet{blanchard_quah89}, the long-run impact of the
demand shock is not transitory in the other non-invertible SVARMA
models whose static shock impact matrix $B$ is estimated using independence
and non-Gaussianity. While the long-run impact of demand shocks is
negative for both non-invertible models, the impact on GNP is always
positive in our approach and the impact on the unemployment rate is
always negative which is in contrast to the results in \citet{GourierouxMR_svarma19}.

\begin{figure}
\includegraphics[scale=0.75]{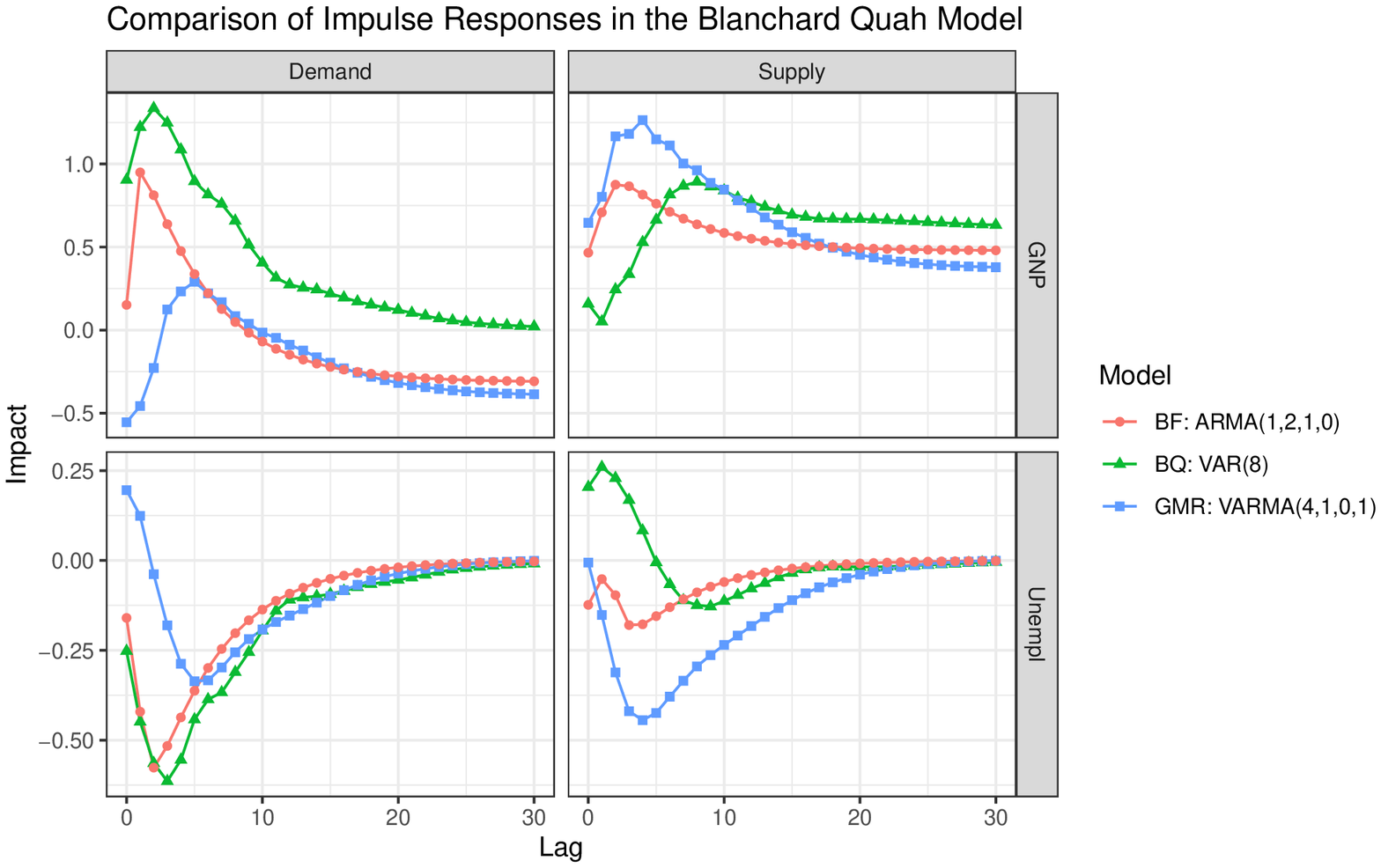}

Figure 1: Comparison of the optimal impulse responses obtained by
different approaches in the literature. In \citet{blanchard_quah89},
the static shock impact matrix is rotated such that the demand shock
is transitory, in the other approaches identification is obtained
by using independence and non-Gaussianity.
\end{figure}

In Figure 2, we illustrate that it is of course possible to identify
the shocks using \citet{blanchard_quah89}'s identification scheme
for the static shock transmission matrix which requires that the long-run
impact on GNP be transitory. In this case, the responses of GNP and
the unemployment rate to the supply shock are more aligned with the
ones in \citet{blanchard_quah89}. One significant difference is that
the response of unemployment to the supply shock is always positive.

\begin{figure}

\includegraphics[scale=0.75]{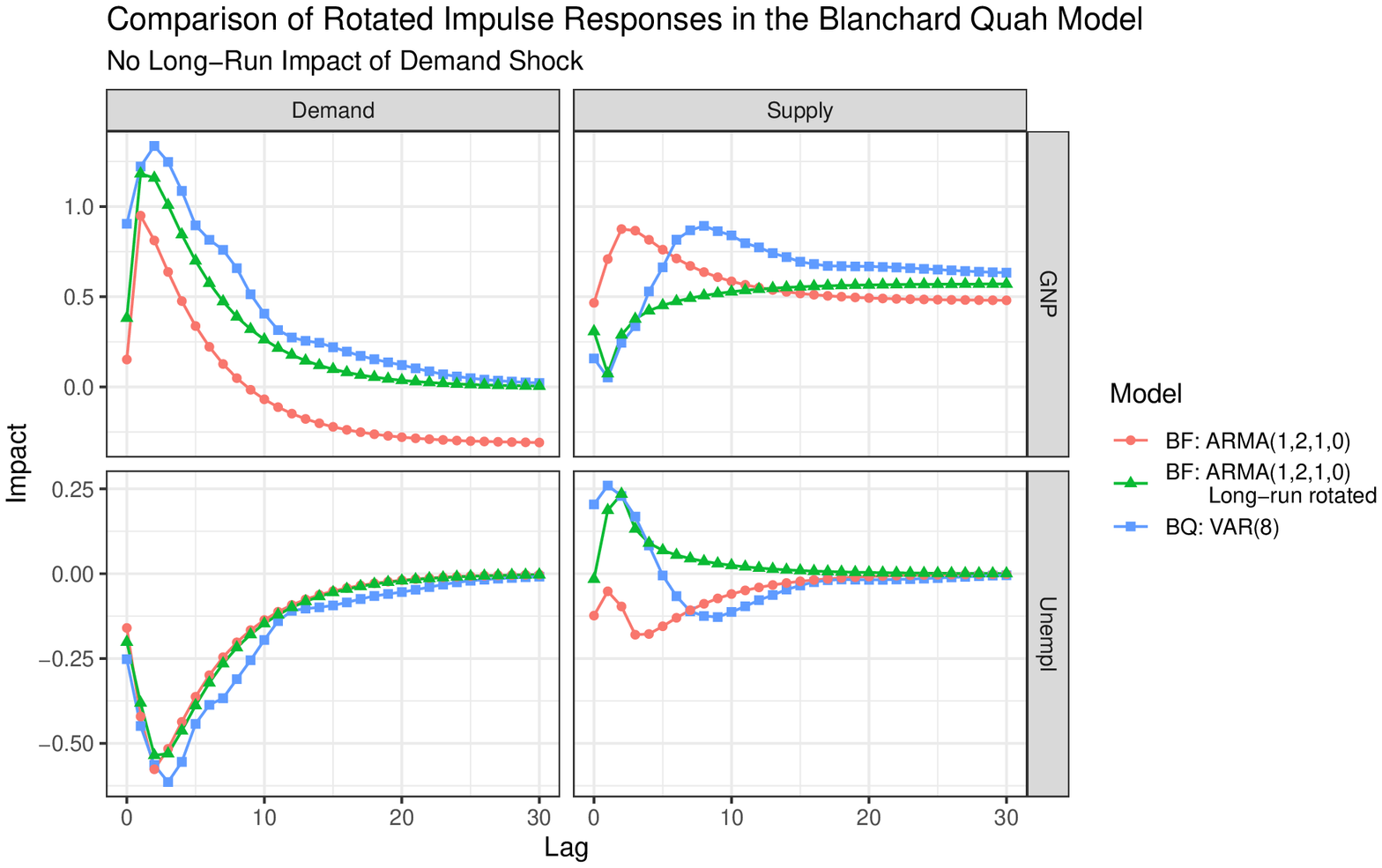}

Figure 2: We compare our optimal model for two different static shock
impact matrices to the one in \citet{blanchard_quah89}. The first
static shock impact matrix corresponds to the optimal one obtained
from the output of ML optimisation, whereas in the second one we rotated
the impulse response function such that the long-run impact of the
demand shock is zero.

\end{figure}

Focusing on the sensitivity of the BIC values for different integer-valued
parameters, we note that these values are quite similar for the dataset
analysed in \citet{blanchard_quah89} and that this is not the case
for different datasets. This suggests that the additional insights
the SVARMA-WHF method provides (e.g. the possibility to estimate each
fundamentalness separately) may be useful for investigating the informativeness
of the data with regards to non-invertibility. Plotting the BIC values
for different $(p,q)$ in dependence of the number of MA roots in
the unit circle, we see that these values are quite similar.

\begin{figure}
\includegraphics[scale=0.4]{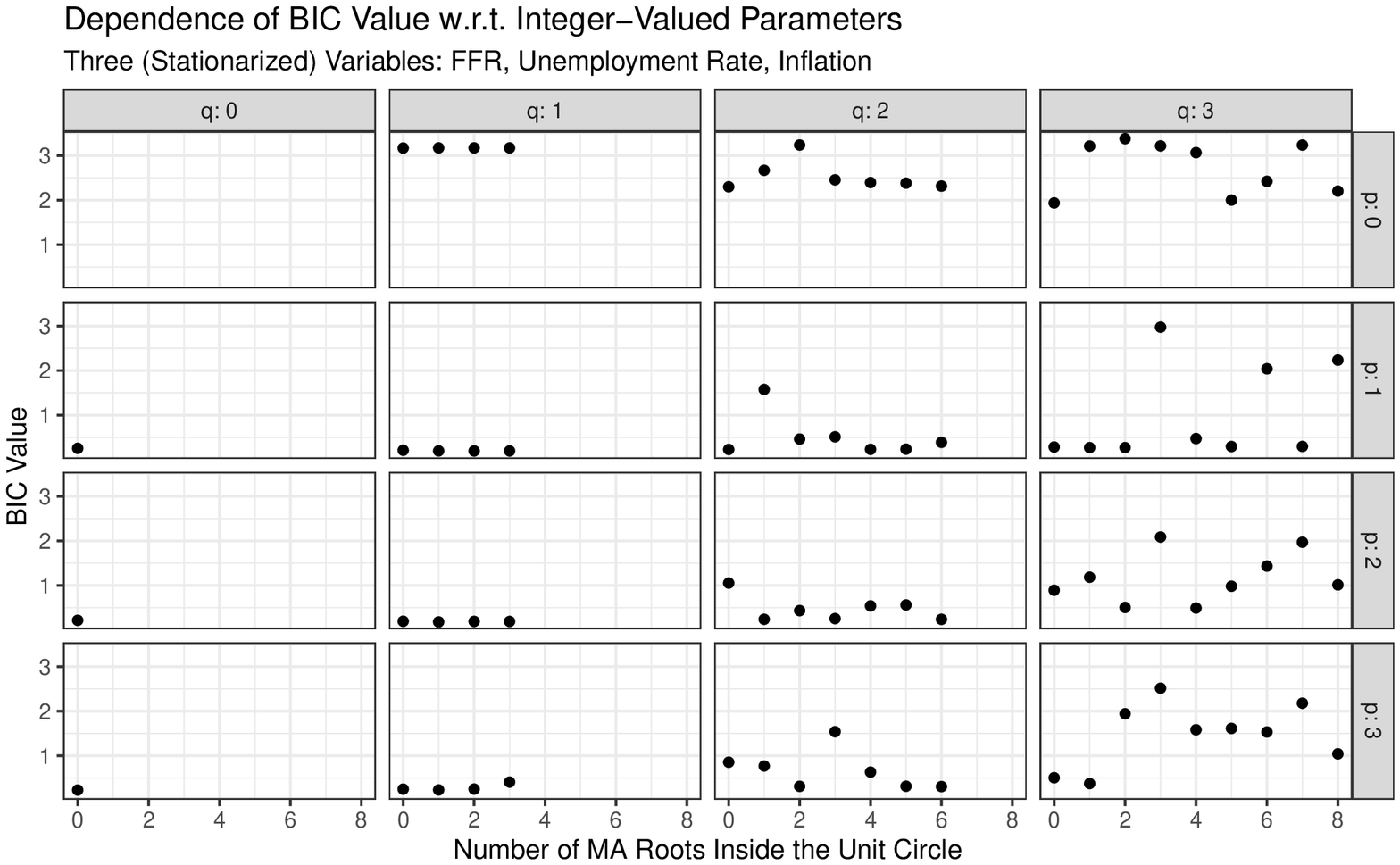}\includegraphics[scale=0.4]{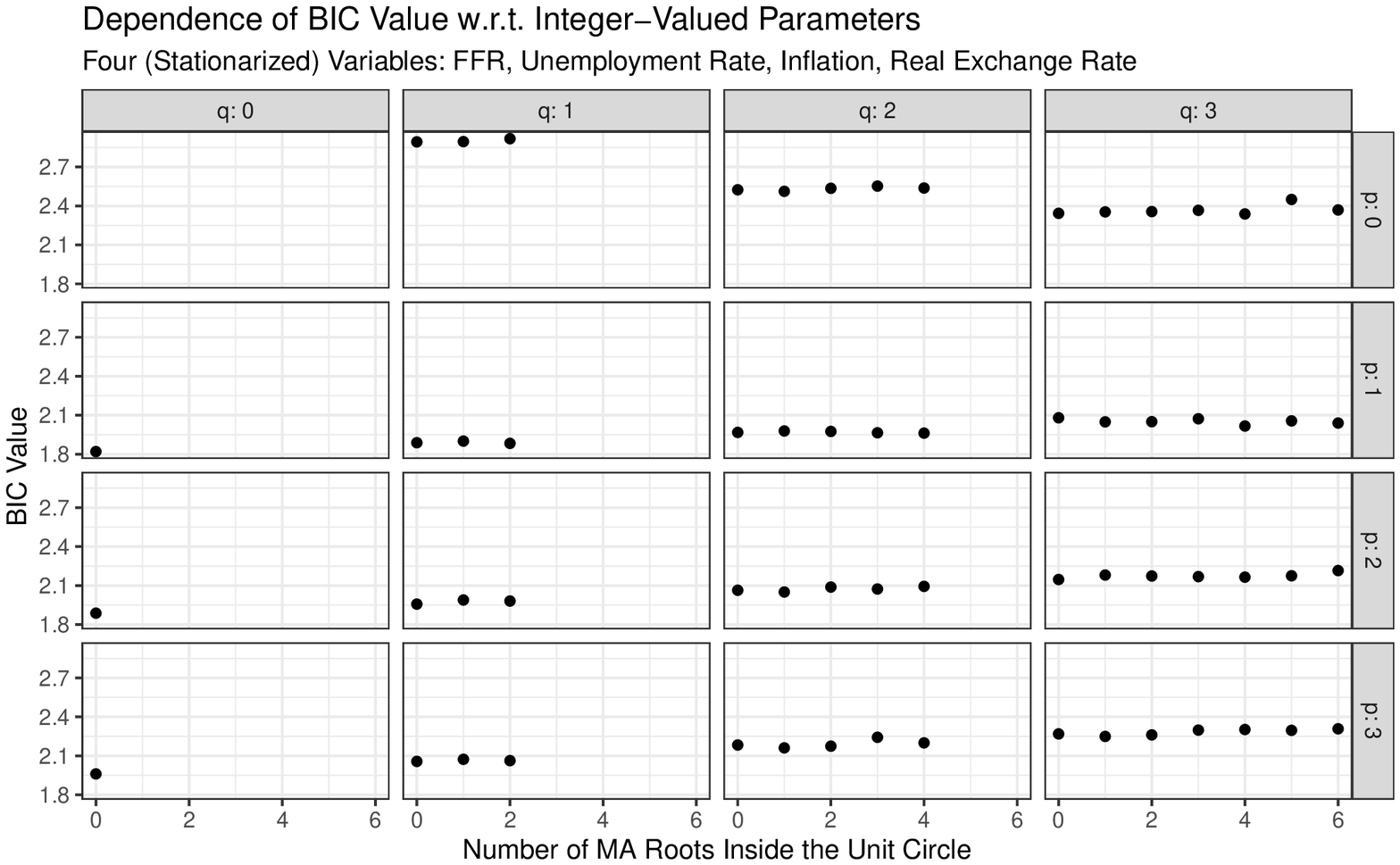}

Figure 3: We compare the dependence of BIC values with respect to
different orders $(p,q)$ and number of unstable roots inside the
unit circle for two standard macroeconomic models. Including additionally
the real exchange rate has the effect that non-invertibility disappears
in the sense that the best model is in this case the one with no MA
part.
\end{figure}

We also considered two standard macroeconomic datasets to illustrate
the dependence of BIC values on integer-valued parameters, in particular
the number of MA zeros inside the unit circle. Unlike the dataset
used in the Blanchard and Quah model, there is a distinct difference
regarding integer-valued parameters and in particular the optimal
number of MA zeros (inside the unit circle) as we illustrate in Figure
3. When the real exchange rate is included in addition to the Federal
Funds Rate, the unemployment rate, and inflation, models without MA
part have a significantly better BIC value.

The SVARMA WHF method presented here suggests itself for likelihood-based
testing procedures. The developments in this article make thus usually
imposed identification restrictions on the zero location of the MA
polynomial testable. Regarding the static shock transmission matrix,
similar remarks to the ones in \citet{LMS_svarIdent16} apply.

More detail regarding the implementation can be found in the documentation
of the associated R-package which can be downloaded from \href{https://github.com/bfunovits/}{https://github.com/bfunovits/}.
Since every non-fundamentalness regime (for all AR and MA orders)
is estimated separately, the SVARMA-WHF approach is ``embarrassingly
parallel'', i.e. little or no effort is needed to separate the problem
into a number of parallel tasks. For $\left(p,q\right)\in\left\{ 0,\ldots,8\right\} ^{2}$
and all fundamentalness regimes, this leads for the Blanchard and
Quah model to 730 different optimisation problems and for the standard
macroeconomic dataset with 4 variables to 1380 different optimisation
problems (which are separated in blocks of 10) incurring a total runtime
of less than 10 minutes. The approaches in \citet{GourierouxMR_svarma19,velasco2020identification}
based on Blaschke matrices return for each $\left(p,q\right)$ only
one value. For $n=4$ and $q=8$, as in the standard macroeconomic
dataset mentioned above, this corresponds for any model with $q=8$
a maximum of $2^{nq}$ (more than 4 billion) optimisations if all
MA zeros are real. In the approach by \citet{velasco2020identification},
this issue is aggravated by the fact that evaluation of higher order
periodograms is extremely costly (requiring up to $\mathcal{O}\left(T^{3}\right)$
calculations) and that the estimator based on higher order cumulant
spectra serves only as an initial estimate for an efficient Newton-Raphson
step minimising the score vector of his concentrated loss functions.

\section{Acknowledgements}

Financial support by the Research Funds of the University of Helsinki
as well as by funds of the Oesterreichische Nationalbank (Austrian
Central Bank, Anniversary Fund, project number: 17646) is gratefully
acknowledged. For computations, the \textit{Finnish Grid and Cloud
Infrastructure} with persistent identifier \textit{urn:nbn:fi:research-infras-2016072533}
was used. Juho Koistinen, Mika Meitz, Markku Lanne, and Wolfgang Scherrer
provided helpful comments on various versions of this article.

\section{Conclusion and Future Research}

In this article, we solve the problem of identifiability and estimation
of causal, possibly non-invertible SVARMA models \eqref{eq:system}
driven by independent and non-Gaussian shocks using a novel parametrisation.
The observation of Prof. Hannan in his Econometric Theory interview
\citep{pagan85_hannanETinterview} applies also here: ``It was really
quite simple once you recognise what the underlying mathematical technique
is. {[}...{]} in the case of that identification problem {[}for causal
and invertible VARMA models{]}, once you recognize the issue as having
to get rid of these indeterminacies, then it can be done very quickly.''

We get rid of the indeterminacies associated with zeros and poles
at infinity of the MA matrix polynomial by using the WHF which ensures
that $f_{0}$ is non-singular. The unique choice of a WHF for a given
MA polynomial, described in Theorem \ref{thm:whf} and Corollary \ref{cor:unique_whf},
is a main ingredient for proving identifiability of the model. The
number of determinantal roots of the MA polynomial is parametrised
by the partial indices of the Wiener-Hopf factorisation of the MA
polynomial. We do not need to assume that the MA root location and
the static shock transmission matrix are identifiable but obtain identifiability
as a consequence of independent shocks and non-Gaussianity. There
are thus no implicit assumptions regarding information symmetry between
outside observers and economic agents in our impulse response analysis.
The structural insights are used for devising a computationally feasible
maximum likelihood estimator which is implemented in the open-source
software R and downloadable from \href{https://github.com/bfunovits/}{https://github.com/bfunovits/}.

An important open question concerns the shock densities, which are
assumed to be known up to a scaling parameter. The estimates obtained
here could serve as initial estimates for an adaptive estimation procedure
along the lines of \citet{kreiss87adaptive,Gassiat90,gassiat1993}.
The theoretical results in Theorem \ref{thm:whf} and Corollary \ref{cor:unique_whf}
can also be used for analysing possibly non-causal VAR models, and
possibly non-causal and possibly non-invertible VARMA models.


\begin{thebibliography}{98}
	\providecommand{\natexlab}[1]{#1}
	\providecommand{\url}[1]{\texttt{#1}}
	\expandafter\ifx\csname urlstyle\endcsname\relax
	\providecommand{\doi}[1]{doi: #1}\else
	\providecommand{\doi}{doi: \begingroup \urlstyle{rm}\Url}\fi
	
	\bibitem[Al-Sadoon(2018)]{AlSadoon18_ET_lrem}
	Majid~M. Al-Sadoon.
	\newblock The linear systems approach to linear rational expectations models.
	\newblock \emph{Econometric Theory}, 34\penalty0 (3):\penalty0 628--658, 2018.
	\newblock \doi{10.1017/S0266466617000160}.
	
	\bibitem[Al-Sadoon and Zwiernik(2019)]{alsadoon2019identification}
	Majid~M. Al-Sadoon and Piotr Zwiernik.
	\newblock The identification problem for linear rational expectations models,
	2019.
	
	\bibitem[Alessi et~al.(2011)Alessi, Barigozzi, and Capasso]{Alessi11noninv}
	Lucia Alessi, Matteo Barigozzi, and Marco Capasso.
	\newblock {Non-Fundamentalness in Structural Econometric Models: A Review}.
	\newblock \emph{International Statistical Review / Revue Internationale de
		Statistique}, 79\penalty0 (1):\penalty0 16--47, 2011.
	\newblock \doi{10.2307/41306163}.
	
	\bibitem[Alpay and Gohberg(1988)]{alpgohberg88}
	Daniel Alpay and Israel Gohberg.
	\newblock \emph{{Topics in Interpolation Theory of Rational Matrix-valued
			Functions}}, chapter {Unitary Rational Matrix Functions}, pages 175--222.
	\newblock 1988.
	
	\bibitem[Anderson and Moore(2005)]{AndMoo05}
	Brian~D.O. Anderson and John~B. Moore.
	\newblock \emph{Optimal filtering}.
	\newblock Dover Publications, New York, 2005.
	
	\bibitem[Andrews(1987)]{andrews87ulln}
	Donald W.~K. Andrews.
	\newblock Consistency in nonlinear econometric models: A generic uniform law of
	large numbers.
	\newblock \emph{Econometrica}, 55\penalty0 (6):\penalty0 1465--1471, 1987.
	\newblock URL \url{http://www.jstor.org/stable/1913568}.
	
	\bibitem[Athanasopoulos and Vahid(2008{\natexlab{a}})]{AthVahid08}
	George Athanasopoulos and Farshid Vahid.
	\newblock Varma versus var for macroeconomic forecasting.
	\newblock \emph{Journal of Business \& Economic Statistics}, 26\penalty0
	(2):\penalty0 237--252, 2008{\natexlab{a}}.
	\newblock \doi{10.1198/073500107000000313}.
	
	\bibitem[Athanasopoulos and Vahid(2008{\natexlab{b}})]{AthVahid_JTSA_08}
	George Athanasopoulos and Farshid Vahid.
	\newblock A complete varma modelling methodology based on scalar components.
	\newblock \emph{Journal of Time Series Analysis}, 29\penalty0 (3):\penalty0
	533--554, 2008{\natexlab{b}}.
	\newblock \doi{10.1111/j.1467-9892.2007.00568.x}.
	
	\bibitem[Baggio and Ferante(2016)]{BaggioFerrante16}
	Giacomo Baggio and Augusto Ferante.
	\newblock On the factorization of rational discrete-time spectral densities.
	\newblock \emph{IEEE Transactions on Automatic Control}, 61\penalty0
	(4):\penalty0 969--981, 2016.
	
	\bibitem[{Baggio} and
	{Ferrante}(2019)]{BaggioFerrante19_parametrization_phasefunction}
	Giacomo {Baggio} and Augusto {Ferrante}.
	\newblock Parametrization of minimal spectral factors of discrete-time rational
	spectral densities.
	\newblock \emph{IEEE Transactions on Automatic Control}, 64\penalty0
	(1):\penalty0 396--403, Jan 2019.
	\newblock \doi{10.1109/TAC.2018.2829474}.
	
	\bibitem[Basilio(2002)]{basilio02}
	Joao~Carlos Basilio.
	\newblock Inversion of polynomial matrices via state-space.
	\newblock \emph{Linear Algebra and its Applications}, 357\penalty0
	(1):\penalty0 259 -- 271, 2002.
	\newblock ISSN 0024-3795.
	\newblock \doi{10.1016/S0024-3795(02)00418-4}.
	\newblock URL
	\url{http://www.sciencedirect.com/science/article/pii/S0024379502004184}.
	
	\bibitem[Blanchard and Quah(1989)]{blanchard_quah89}
	Olivier~J Blanchard and Danny Quah.
	\newblock {The Dynamic Effects of Aggregate Demand and Supply Disturbances}.
	\newblock \emph{American Economic Review}, 79\penalty0 (4):\penalty0 655--673,
	1989.
	
	\bibitem[B\"ottcher and Grudsky(2005)]{BoettcherGrudsky05_bandedtoeplitz}
	Albrecht B\"ottcher and Sergei~M. Grudsky.
	\newblock \emph{{Spectral Properties of Banded Toeplitz Matrices}}.
	\newblock SIAM, 2005.
	
	\bibitem[Brillinger(1981)]{Brillinger75}
	David~R. Brillinger.
	\newblock \emph{Time Series: Data Analysis and Theory}.
	\newblock Holden Day, San Francisco, 1981.
	
	\bibitem[Brockwell and Davis(1987)]{BrockwellDavis87}
	P.~J. Brockwell and R.~A. Davis.
	\newblock \emph{Time Series: Data Theory and Methods}.
	\newblock Springer, New York, 1987.
	
	\bibitem[Burnham and Anderson(2004)]{burnhamanderson04modelselection}
	Kenneth~P Burnham and David~R Anderson.
	\newblock Multimodel inference: understanding aic and bic in model selection.
	\newblock \emph{Sociological methods \& research}, 33\penalty0 (2):\penalty0
	261--304, 2004.
	\newblock \doi{10.1007/b97636}.
	
	\bibitem[Chan and Ho(2004)]{chanho04}
	Kung-Sik Chan and Lop-Hing Ho.
	\newblock {On the Unique Representation of non-Gaussian Multivariate Linear
		Processes}.
	\newblock Technical report, Department of Statistics \& Actuarial Science,
	University of Iowa., 2004.
	\newblock URL
	\url{https://pdfs.semanticscholar.org/fd93/194a3fd280de596ef5135aa7954eab5e51a1.pdf}.
	
	\bibitem[Chan et~al.(2006)Chan, Ho, and Tong]{chanHoTong06}
	Kung-Sik Chan, Lop-Hing Ho, and Howell Tong.
	\newblock {A Note on Time-Reversibility of Multivariate Linear Processes}.
	\newblock \emph{Biometrika}, 93:\penalty0 221--227, 2006.
	\newblock \doi{10.1093/biomet/93.1.221}.
	
	\bibitem[Chen and Bickel(2005)]{ChenBickel05}
	Aiyou Chen and Peter~J. Bickel.
	\newblock Consistent independent component analysis and prewhitening.
	\newblock \emph{IEEE Transactions on Signal Processing}, 53:\penalty0
	3625--3632, 2005.
	\newblock \doi{10.1109/TSP.2005.855098}.
	
	\bibitem[Cheng(1990)]{cheng1990}
	Qiansheng Cheng.
	\newblock Maximum standardized cumulant deconvolution of non-gaussian linear
	processes.
	\newblock \emph{Ann. Statist.}, 18\penalty0 (4):\penalty0 1774--1783, 12 1990.
	\newblock \doi{10.1214/aos/1176347877}.
	\newblock URL \url{https://doi.org/10.1214/aos/1176347877}.
	
	\bibitem[Cheng(1992)]{cheng1992}
	Qiansheng Cheng.
	\newblock On the unique representation of non-gaussian linear processes.
	\newblock \emph{Ann. Statist.}, 20\penalty0 (2):\penalty0 1143--1145, 06 1992.
	\newblock \doi{10.1214/aos/1176348677}.
	\newblock URL \url{https://doi.org/10.1214/aos/1176348677}.
	
	\bibitem[Claeskens and Hjort(2008)]{claeskenshjort08model}
	Gerda Claeskens and Nils~Lid Hjort.
	\newblock Model selection and model averaging.
	\newblock \emph{Cambridge Books}, 2008.
	
	\bibitem[Clancey and Gohberg(1981)]{clanceygohberg81}
	Kevin~F. Clancey and Israel Gohberg.
	\newblock \emph{Factorization of Matrix Functions and Singular Integral
		Operators}.
	\newblock Springer Basel AG, 1981.
	
	\bibitem[Dahlhaus and P{\"o}tscher(1989)]{dahlhauspoet89}
	R~Dahlhaus and B.M P{\"o}tscher.
	\newblock Convergence results for maximum likelihood type estimators in
	multivariable arma models ii.
	\newblock \emph{Journal of Multivariate Analysis}, 30\penalty0 (2):\penalty0
	241 -- 244, 1989.
	\newblock ISSN 0047-259X.
	\newblock \doi{https://doi.org/10.1016/0047-259X(89)90037-7}.
	\newblock URL
	\url{http://www.sciencedirect.com/science/article/pii/0047259X89900377}.
	
	\bibitem[Davis(2015)]{davis16sgt_pkg}
	Carter Davis.
	\newblock \emph{sgt: Skewed Generalized T Distribution Tree}, 2015.
	\newblock URL \url{https://CRAN.R-project.org/package=sgt}.
	\newblock R package version 2.0.
	
	\bibitem[Davis and Song(2010)]{davis10}
	R~Davis and L~Song.
	\newblock Noncausal vector ar processes with application to financial time
	series.
	\newblock \emph{Unpublished manuscript, Columbia University, New York}, 2010.
	
	\bibitem[Deistler(1975)]{deistler75}
	Manfred Deistler.
	\newblock {z-Transform and Identification of Linear Econometric Models with
		Autocorrelated Errors}.
	\newblock \emph{Metrika}, 22:\penalty0 13--25, 1975.
	
	\bibitem[Deistler(1983)]{deistler83}
	Manfred Deistler.
	\newblock {The Properties of the Parameterization of ARMAX Systems and Their
		Relevance for Structural Estimation and Dynamic Specification}.
	\newblock \emph{Econometrica}, 51\penalty0 (4):\penalty0 1187--1207, July 1983.
	\newblock URL \url{http://www.jstor.org/stable/1912058}.
	
	\bibitem[Deistler and P\"otscher(1984)]{DeistlerPoetscher84}
	Manfred Deistler and Benedikt~M. P\"otscher.
	\newblock {The Behaviour of the Likelihood Function for ARMA Models}.
	\newblock \emph{Advances in Applied Probability}, 16\penalty0 (4):\penalty0
	843--866, December 1984.
	
	\bibitem[Deistler and Seifert(1978)]{DeistlerSeifert78}
	Manfred Deistler and Hans-G\"unther Seifert.
	\newblock {Identifiability and Consistent Estimability in Econometric Models}.
	\newblock \emph{Econometrica}, 46\penalty0 (6):\penalty0 969--980, July 1978.
	\newblock URL \url{http://www.jstor.org/stable/1909759}.
	
	\bibitem[Findley(1986)]{findley86}
	David~F. Findley.
	\newblock The uniqueness of moving average representations with independent and
	identically distributed random variables for non-gaussian stationary time
	series.
	\newblock \emph{Biometrika}, 73\penalty0 (2):\penalty0 520--521, 08 1986.
	\newblock \doi{10.1093/biomet/73.2.520}.
	
	\bibitem[Findley(1990)]{findley90}
	David~F. Findley.
	\newblock Amendments and corrections: The uniqueness of moving average
	representations with independent and identically distributed random variables
	for non-gaussian stationary time series.
	\newblock \emph{Biometrika}, 77\penalty0 (1):\penalty0 235--235, 1990.
	\newblock URL \url{http://www.jstor.org/stable/2336071}.
	
	\bibitem[Forni et~al.(2019)Forni, Gambetti, and Sala]{forniGambetSala19_noninv}
	Mario Forni, Luca Gambetti, and Luca Sala.
	\newblock {Structural VARs and noninvertible macroeconomic models}.
	\newblock \emph{Journal of Applied Econometrics}, 34\penalty0 (2):\penalty0
	221--246, 2019.
	\newblock \doi{10.1002/jae.2665}.
	\newblock URL \url{https://onlinelibrary.wiley.com/doi/abs/10.1002/jae.2665}.
	
	\bibitem[Funovits(2019)]{funo_allpass}
	Bernd Funovits.
	\newblock Semi-parametric estimation of multivariate possibly non-causal and
	possibly non-invertible time series models.
	\newblock \emph{Manuscript, University of Helsinki}, 2019.
	
	\bibitem[Funovits(2020)]{funovits2020gmr_comment}
	Bernd Funovits.
	\newblock Comment on gouri\'eroux, monfort, renne (2019): Identification and
	estimation in non-fundamental structural varma models, 2020.
	
	\bibitem[Gantmacher(1959)]{Gant1}
	Felix~R. Gantmacher.
	\newblock \emph{The Theory of Matrices}, volume~1.
	\newblock AMS Chelsea Publishing, 1959.
	
	\bibitem[Gassiat(1993)]{gassiat1993}
	E.~Gassiat.
	\newblock Adaptive estimation in noncausal stationary ar processes.
	\newblock \emph{Ann. Statist.}, 21\penalty0 (4):\penalty0 2022--2042, 12 1993.
	\newblock \doi{10.1214/aos/1176349408}.
	\newblock URL \url{https://doi.org/10.1214/aos/1176349408}.
	
	\bibitem[Gassiat(1990)]{Gassiat90}
	\'Elisabeth Gassiat.
	\newblock Estimation semi-param\'etrique d'un mod\`ele autor\'egressif
	stationnaire multiindice non n\'ecessairement causal.
	\newblock \emph{Annales de l'I.H.P. Probabilit\'es et statistiques},
	26\penalty0 (1):\penalty0 181--205, 1990.
	\newblock URL \url{www.numdam.org/item/AIHPB_1990__26_1_181_0/}.
	
	\bibitem[Gohberg and Feldman(1974)]{gohberg_feldman81}
	Israel Gohberg and Izrail~Aronovich Feldman.
	\newblock \emph{Convolution equations and projection methods for their
		solution}, volume~41.
	\newblock American Mathematical Society, translations of mathematical
	monographs edition, 1974.
	
	\bibitem[Gohberg and Krein(1960)]{GohbergKrein60}
	Israel Gohberg and M.~G. Krein.
	\newblock {Systems of integral equations on a half-line with kernel
		dependingupon the dierence of the arguments}.
	\newblock 14\penalty0 (2):\penalty0 217--287, 1960.
	
	\bibitem[Gohberg et~al.(2003)Gohberg, Kaashoek, and
	Spitkovsky]{gohkaaspit03_summerschool}
	Israel Gohberg, Marinus.~A. Kaashoek, and Ilya~M. Spitkovsky.
	\newblock {An Overview of Matrix Factorization Theory and Operator
		Applications}.
	\newblock In Israel Gohberg, Nenad Manojlovic, and Ant{\'o}nio~Ferreira dos
	Santos, editors, \emph{Factorization and Integrable Systems}, pages 1--102,
	Basel, 2003. Birkh{\"a}user Basel.
	\newblock ISBN 978-3-0348-8003-9.
	
	\bibitem[Gohberg et~al.(2009)Gohberg, Lancaster, and
	Rodman]{GohbergLancasterRodman09}
	Israel Gohberg, Peter Lancaster, and Leiba Rodman.
	\newblock \emph{{Matrix Polynomials}}.
	\newblock SIAM, Philadelphia, 2009.
	
	\bibitem[Gouri\'eroux and Jasiak(2017)]{goujas17_noncausal_semiparam}
	Christian Gouri\'eroux and Joann Jasiak.
	\newblock Noncausal vector autoregressive process: Representation,
	identification and semi-parametric estimation.
	\newblock \emph{Journal of Econometrics}, 200\penalty0 (1):\penalty0 118--134,
	2017.
	\newblock \doi{http://dx.doi.org/10.1016/j.jeconom.2017.01.011}.
	
	\bibitem[Gouri\'eroux et~al.(2017)Gouri\'eroux, Monfort, and
	Renne]{GourierouxZakoianRenne17}
	Christian Gouri\'eroux, Alain Monfort, and Jean-Paul Renne.
	\newblock {Statistical inference for independent component analysis:
		Application to structural VAR models}.
	\newblock \emph{Journal of Econometrics}, 196:\penalty0 111--126, 2017.
	\newblock \doi{10.1016/j.jeconom.2016.09.007}.
	
	\bibitem[Gouri\'eroux et~al.(2019)Gouri\'eroux, Monfort, and
	Renne]{GourierouxMR_svarma19}
	Christian Gouri\'eroux, Alain Monfort, and Jean-Paul Renne.
	\newblock {Identification and Estimation in Non-Fundamental Structural VARMA
		Models}.
	\newblock \emph{Review of Economic Studies}, pages 1--39, 2019.
	\newblock \doi{10.1093/restud/rdz028}.
	
	\bibitem[Hallin and Mehta(2015)]{HallinMehta15}
	Marc Hallin and Chintan Mehta.
	\newblock {R-Estimation for Asymmetric Independent Component Analysis}.
	\newblock \emph{Journal of the American Statistical Association}, 110\penalty0
	(509):\penalty0 218--232, 2015.
	\newblock \doi{10.1080/01621459.2014.909316}.
	
	\bibitem[Hannan(1970)]{Hannan70}
	Edward~J. Hannan.
	\newblock \emph{Multiple Time Series}.
	\newblock Wiley, 1970.
	
	\bibitem[Hannan(1971)]{Hannan71}
	Edward~J. Hannan.
	\newblock The identification problem for multiple equation systems with moving
	average errors.
	\newblock \emph{Econometrica}, 39\penalty0 (5):\penalty0 751--765, September
	1971.
	\newblock \doi{10.2307/1909577}.
	
	\bibitem[Hannan and Deistler(2012)]{HannanDeistler12}
	Edward~J. Hannan and Manfred Deistler.
	\newblock \emph{The Statistical Theory of Linear Systems}.
	\newblock SIAM Classics in Applied Mathematics, Philadelphia, 2012.
	
	\bibitem[Hannan and Poskitt(1988)]{hannanposkitt1988}
	Edward~J. Hannan and Donald~S. Poskitt.
	\newblock {Unit Canonical Correlations between Future and Past}.
	\newblock \emph{The Annals of Statistics}, 16\penalty0 (2):\penalty0 784--790,
	June 1988.
	\newblock \doi{10.1214/aos/1176350836}.
	
	\bibitem[Hansen and Sargent(1991)]{HansenSargent91_2difficulties}
	Lars~Peter Hansen and Thomas~J. Sargent.
	\newblock \emph{{Rational Expectations Econometrics}}, chapter {Two
		Difficulties in Interpreting Vector Autoregressions}, pages 77--120.
	\newblock Westview Press, 1991.
	
	\bibitem[Harville(1997)]{Harville97}
	David~A. Harville.
	\newblock \emph{{Matrix Algebra From a Statistician's Perspective}}.
	\newblock Springer, 1997.
	
	\bibitem[Hinrichsen and Pritchard(2005)]{HinrichsenPritchard00}
	Diederich Hinrichsen and Anthony~J. Pritchard.
	\newblock \emph{Mathematical Systems Theory I - Modelling, State Space
		Analysis, Stability and Robustness}.
	\newblock Springer-Verlag, Berlin, Heidelberg, 2005.
	
	\bibitem[Hyv\"arinen et~al.(2001)Hyv\"arinen, Karhunen, and Oja]{Hyvarinen01}
	Aapo Hyv\"arinen, Juha Karhunen, and Erkki Oja.
	\newblock \emph{Independent Component Analysis}.
	\newblock John Wiley \& Sons, 2001.
	
	\bibitem[Ilmonen and Paindaveine(2011)]{IlmonenPaindaveine11}
	Pauliina Ilmonen and Davy Paindaveine.
	\newblock Semiparametrically efficient inference based on signed ranks in
	symmetric independent component models.
	\newblock \emph{Annals of Statistics}, 39\penalty0 (5):\penalty0 2448--2476,
	2011.
	\newblock \doi{10.1214/11-AOS906}.
	
	\bibitem[Jammalamadaka et~al.(2006)Jammalamadaka, Rao, and
	Terdik]{Jammalamadaka_rao_terdik06}
	S.~Rao Jammalamadaka, T.~Subba Rao, and Gy\"orgy Terdik.
	\newblock Higher order cumulants of random vectors and applications to
	statistical inference and time series.
	\newblock \emph{Sankhya: The Indian Journal of Statistics}, 68\penalty0
	(2):\penalty0 326--356, 2006.
	\newblock \doi{10.2307/25053499}.
	
	\bibitem[Kagan et~al.(1973)Kagan, Linnik, and Rao]{Kagan73}
	Abram~M. Kagan, Yuri~V. Linnik, and Calyampudi~R. Rao.
	\newblock \emph{{Characterization Problems in Mathematical Statistics}}.
	\newblock John Wiley \& Sons, 1973.
	
	\bibitem[Kailath(1980)]{Kailath1980}
	Thomas Kailath.
	\newblock \emph{{Linear Systems}}.
	\newblock Prentice Hall, Englewood Cliffs, N.J., 1980.
	
	\bibitem[Kilian and L\"utkepohl(2017)]{KilianLut17}
	Lutz Kilian and Helmut L\"utkepohl.
	\newblock \emph{Structural Vector Autoregressive Analysis}.
	\newblock Cambridge University Press, 2017.
	\newblock \doi{10.1017/9781108164818}.
	
	\bibitem[Kreiss(1987)]{kreiss87adaptive}
	Jens-Peter Kreiss.
	\newblock {On adaptive estimation in stationary ARMA processes}.
	\newblock \emph{The Annals of Statistics}, 15\penalty0 (1):\penalty0 112--133,
	1987.
	\newblock URL \url{https://www.jstor.org/stable/2241072}.
	
	\bibitem[Lanne and Saikkonen(2013)]{LanneSaikkonen13}
	Markku Lanne and Pentti Saikkonen.
	\newblock {Noncausal Vector Autoregression}.
	\newblock \emph{Econometric Theory}, 29:\penalty0 447--481, 6 2013.
	\newblock ISSN 1469-4360.
	\newblock \doi{10.1017/S0266466612000448}.
	
	\bibitem[Lanne et~al.(2017)Lanne, Meitz, and Saikkonen]{LMS_svarIdent16}
	Markku Lanne, Mika Meitz, and Pentti Saikkonen.
	\newblock Identification and estimation of non-gaussian structural vector
	autoregressions.
	\newblock \emph{Journal of Econometrics}, 196\penalty0 (2):\penalty0 288 --
	304, 2017.
	\newblock \doi{10.1016/j.jeconom.2016.06.002}.
	
	\bibitem[Leonov and Shiryaev(1959)]{leonov1959method}
	Valerii~P. Leonov and Albert~Nikolaevich Shiryaev.
	\newblock On a method of calculation of semi-invariants.
	\newblock \emph{{Theory of Probability \& its applications}}, 4\penalty0
	(3):\penalty0 319--329, 1959.
	
	\bibitem[Lii and Rosenblatt(1992)]{LiiRosenblatt92}
	Keh-Shin Lii and Murray Rosenblatt.
	\newblock {An approximate maximum likelihood estimation for non-Gaussian
		non-minimum phase moving average processes}.
	\newblock \emph{Journal of Multivariate Analysis}, 43\penalty0 (2):\penalty0
	272--299, 1992.
	\newblock \doi{10.1016/0047-259X(92)90037-G}.
	
	\bibitem[Lii and Rosenblatt(1996)]{LiiRosenblatt96}
	Keh-Shin Lii and Murray Rosenblatt.
	\newblock {Maximum Likelihood Estimation for Nongaussian Nonminimum Phase ARMA
		Sequences}.
	\newblock \emph{Statistica Sinica}, 6\penalty0 (1):\penalty0 1--22, 1996.
	\newblock \doi{10.2307/24305996}.
	\newblock URL \url{http://www.jstor.org/stable/24305996}.
	
	\bibitem[Lippi and Reichlin(1993)]{lippi_reichlin93aer}
	Marco Lippi and Lucrezia Reichlin.
	\newblock {The Dynamic Effects of Aggregate Demand and Supply Disturbances:
		Comment}.
	\newblock \emph{The American Economic Review}, 83\penalty0 (3):\penalty0
	644--652, 1993.
	
	\bibitem[Lippi and Reichlin(1994)]{LippiReichlin94}
	Marco Lippi and Lucrezia Reichlin.
	\newblock {VAR Analysis, Nonfundamental Representations, Blaschke Matrices}.
	\newblock \emph{Journal of Econometrics}, 63\penalty0 (1):\penalty0 307 -- 325,
	1994.
	\newblock ISSN 0304-4076.
	\newblock \doi{http://dx.doi.org/10.1016/0304-4076(93)01570-C}.
	\newblock URL
	\url{http://www.sciencedirect.com/science/article/pii/030440769301570C}.
	
	\bibitem[McCullagh(2018)]{mccullagh2018tensor}
	Peter McCullagh.
	\newblock \emph{Tensor methods in statistics}.
	\newblock Dover Publications, 2018.
	
	\bibitem[Newton(1978)]{newton1978information}
	H.~Joseph Newton.
	\newblock The information matrices of the parameters of multiple mixed time
	series.
	\newblock \emph{Journal of Multivariate Analysis}, 8\penalty0 (2):\penalty0
	317--323, 1978.
	
	\bibitem[Onatski(2006)]{onatski06}
	Alexei Onatski.
	\newblock {Winding number criterion for existence and uniqueness of equilibrium
		in linear rational expectations models}.
	\newblock \emph{Journal of Economic Dynamics and Control}, 30\penalty0
	(2):\penalty0 323--345, February 2006.
	
	\bibitem[Pagan(1985)]{pagan85_hannanETinterview}
	Adrian Pagan.
	\newblock {The ET Interview: Professor E.J. Hannan}.
	\newblock \emph{Econometric Theory}, 1\penalty0 (2):\penalty0 263--290, 1985.
	\newblock \doi{10.1017/S0266466600011178}.
	
	\bibitem[Plagborg-M{{\o}}ller(2019)]{plagbormoller19SVMA}
	Mikkel Plagborg-M{{\o}}ller.
	\newblock Bayesian inference on structural impulse response functions.
	\newblock \emph{Quantitative Economics}, 10\penalty0 (1):\penalty0 145--184,
	2019.
	\newblock \doi{https://doi.org/10.3982/QE926}.
	\newblock URL \url{https://www.onlinelibrary.wiley.com/doi/abs/10.3982/QE926}.
	
	\bibitem[Poskitt(2016)]{Poskitt16}
	Donald~S. Poskitt.
	\newblock Vector autoregressive moving average identification for macroeconomic
	modeling: A new methodology.
	\newblock \emph{Journal of Econometrics}, 192:\penalty0 468--484, June 2016.
	\newblock \doi{10.1016/j.jeconom.2016.02.011}.
	
	\bibitem[Poskitt and Yao(2017)]{PoskittYao17}
	Donald~S. Poskitt and Wenying Yao.
	\newblock Vector autoregressions and macroeconomic modeling: An error taxonomy.
	\newblock \emph{Journal of Business \& Economic Statistics}, 35\penalty0
	(3):\penalty0 407--419, 2017.
	\newblock \doi{10.1080/07350015.2015.1077139}.
	
	\bibitem[P\"otscher(2011)]{poet11}
	Benedikt P\"otscher.
	\newblock Asymptotic properties of m-estimators - lecture notes.
	\newblock \emph{Manuscript, University of Vienna}, 2011.
	
	\bibitem[P\"otscher and Prucha(1997)]{poetpruch97}
	Benedikt P\"otscher and Ingmar Prucha.
	\newblock \emph{Dynamic Nonlinear Econometric Models, Asymptotic Theory}.
	\newblock Springer Berlin, 1997.
	
	\bibitem[P{\"o}tscher(1990)]{poet90armaorder}
	Benedikt~M. P{\"o}tscher.
	\newblock Estimation of autoregressive moving-average order given an infinite
	number of models and approximation of spectral densities.
	\newblock \emph{Journal of Time Series Analysis}, 11\penalty0 (2):\penalty0
	165--179, 1990.
	\newblock \doi{https://doi.org/10.1111/j.1467-9892.1990.tb00049.x}.
	\newblock URL
	\url{https://onlinelibrary.wiley.com/doi/abs/10.1111/j.1467-9892.1990.tb00049.x}.
	
	\bibitem[P{\"o}tscher(1987)]{poet87}
	B.M. P{\"o}tscher.
	\newblock Convergence results for maximum likelihood type estimators in
	multivariable arma models.
	\newblock \emph{Journal of Multivariate Analysis}, 21\penalty0 (1):\penalty0 29
	-- 52, 1987.
	\newblock ISSN 0047-259X.
	\newblock \doi{https://doi.org/10.1016/0047-259X(87)90097-2}.
	\newblock URL
	\url{http://www.sciencedirect.com/science/article/pii/0047259X87900972}.
	
	\bibitem[Raghavan et~al.(2016)Raghavan, Athanasopoulos, and
	Silvapulle]{RaghavanAthSilvapulle16}
	Mala Raghavan, George Athanasopoulos, and Param Silvapulle.
	\newblock {Canadian monetary policy analysis using a structural VARMA model}.
	\newblock \emph{Canadian Journal of Economics}, 49\penalty0 (1):\penalty0
	347--373, 2016.
	\newblock \doi{10.1111/caje.12200}.
	
	\bibitem[Ravenna(2007)]{Ravenna07}
	Federico Ravenna.
	\newblock Vector autoregressions and reduced form representations of \{DSGE\}
	models.
	\newblock \emph{Journal of Monetary Economics}, 54\penalty0 (7):\penalty0 2048
	-- 2064, 2007.
	\newblock ISSN 0304-3932.
	\newblock \doi{10.1016/j.jmoneco.2006.09.002}.
	
	\bibitem[Reinsel(1993)]{Reinsel93}
	G.~Reinsel.
	\newblock \emph{Elements of Multivariate Time Series Analysis}.
	\newblock Springer Series in Statistics, 1993.
	
	\bibitem[Rissanen and Caines(1979)]{rissanencaines79mle}
	J.~Rissanen and P.~E. Caines.
	\newblock The strong consistency of maximum likelihood estimators for arma
	processes.
	\newblock \emph{The Annals of Statistics}, 7\penalty0 (2):\penalty0 297--315,
	1979.
	\newblock \doi{10.2307/2958812}.
	
	\bibitem[Rosenblatt(1985)]{rosenblatt1985}
	Murray Rosenblatt.
	\newblock \emph{Stationary sequences and random fields}.
	\newblock Springer Science \& Business Media, 1985.
	
	\bibitem[Rosenblatt(2000)]{Rosenblatt00}
	Murray Rosenblatt.
	\newblock \emph{{Gaussian and Non-Gaussian Linear Time Series and Random
			Fields}}.
	\newblock Springer, 2000.
	
	\bibitem[Rothenberg(1971)]{Rothenberg71}
	Thomas~J. Rothenberg.
	\newblock {Identification in Parametric Models}.
	\newblock \emph{Econometric Theory}, 39\penalty0 (3):\penalty0 577--591, May
	1971.
	\newblock \doi{10.2307/1913267}.
	
	\bibitem[Rozanov(1967)]{Rozanov67}
	Yuri~A. Rozanov.
	\newblock \emph{{Stationary Random Processes}}.
	\newblock Holden-Day, San Francisco, 1967.
	
	\bibitem[Scherrer and Funovits(2020{\natexlab{a}})]{ScherrerFuno_ratmat}
	Wolfgang Scherrer and Bernd Funovits.
	\newblock \emph{{rationalmatrices: Classes and Methods for Rational Matrices}},
	2020{\natexlab{a}}.
	
	\bibitem[Scherrer and Funovits(2020{\natexlab{b}})]{ScherrerFuno_rldm}
	Wolfgang Scherrer and Bernd Funovits.
	\newblock \emph{{RLDM: A Package for Modeling of Time Series with a Rational
			Spectral Density}}, 2020{\natexlab{b}}.
	
	\bibitem[Scherrer and
	Funovits(2020{\natexlab{c}})]{scherrer_funovits2020allpass}
	Wolfgang Scherrer and Bernd Funovits.
	\newblock All-pass functions for mirroring pairs of complex-conjugated roots of
	rational matrix functions, 2020{\natexlab{c}}.
	
	\bibitem[Seber(2008)]{seber08}
	George A.~F. Seber.
	\newblock \emph{{A Matrix Handbook for Statisticians}}.
	\newblock John Wiley \& Sons, 2008.
	
	\bibitem[Sims(1980)]{Sims80}
	Christopher~A. Sims.
	\newblock Macroeconomics and reality.
	\newblock \emph{Econometrica}, 48\penalty0 (1):\penalty0 1--48, 1980.
	\newblock \doi{10.2307/1912017}.
	\newblock URL \url{http://www.jstor.org/stable/1912017}.
	
	\bibitem[Sims and Zha(2006)]{simsZha06monpol_recessioni}
	Christopher~A. Sims and Tao Zha.
	\newblock Does monetary policy generate recessions?
	\newblock \emph{Macroeconomic Dynamics}, 10\penalty0 (2):\penalty0 231--272,
	2006.
	\newblock \doi{10.1017/S136510050605019X}.
	
	\bibitem[Theodossiou(1998)]{theodossiou98sgt}
	Panayiotis Theodossiou.
	\newblock Financial data and the skewed generalized t distribution.
	\newblock \emph{Management Science}, 44\penalty0 (12):\penalty0 1650--1661,
	1998.
	\newblock \doi{10.2307/2634700}.
	
	\bibitem[Velasco(2020)]{velasco2020identification}
	Carlos Velasco.
	\newblock Identification and estimation of structural varma models using higher
	order dynamics, 2020.
	
	\bibitem[Velasco and Lobato(2018)]{VelascoLobato18}
	Carlos Velasco and Ignacio~N. Lobato.
	\newblock Frequency domain minimum distance inference for possibly
	noninvertible and noncausal arma models.
	\newblock \emph{Ann. Statist.}, 46\penalty0 (2):\penalty0 555--579, 04 2018.
	\newblock \doi{10.1214/17-AOS1560}.
	\newblock URL \url{https://doi.org/10.1214/17-AOS1560}.
	
	\bibitem[Wolovich(1974)]{wolovich74}
	William~A. Wolovich.
	\newblock \emph{{Linear Multivariable Systems}}.
	\newblock Springer Verlag, third edition, 1974.
	
	\bibitem[Yao et~al.(2017)Yao, Kam, and Vahid]{YaoKamVahid17weakVARMA}
	Wenying Yao, Timothy Kam, and Farshid Vahid.
	\newblock On weak identification in structural varma models.
	\newblock \emph{Economics Letters}, 156:\penalty0 1--6, 2017.
	\newblock ISSN 0165-1765.
	\newblock \doi{10.1016/j.econlet.2017.03.035}.
	\newblock URL
	\url{https://www.sciencedirect.com/science/article/pii/S0165176517301374}.
	
	\bibitem[Zurbenko(1986)]{zurb86}
	Igor~G. Zurbenko.
	\newblock \emph{The spectral analysis of time series}.
	\newblock Amsterdam ; New York : North-Holland, Elsevier Science Pub. Co, 1986.
	\newblock ISBN 0444876073.
	
\end{thebibliography}
\end{document}